\documentclass[superscriptaddress,aps,prd,showpacs, twocolumn]{revtex4}
\usepackage{graphicx}
\usepackage{amsmath}
\usepackage{bm}
\usepackage{color}

\def\be{\begin{equation}}
\def\ee{\end{equation}}
\def\bea{\begin{eqnarray}}
\def\eea{\end{eqnarray}}

\begin{document}

\title{Arbitrary scalar field and quintessence cosmological models}
\author{Tiberiu Harko}
\email{t.harko@ucl.ac.uk}
\affiliation{Department of Mathematics, University College London, Gower Street, London
WC1E 6BT, United Kingdom}
\author{Francisco S. N. Lobo}
\email{flobo@cii.fc.ul.pt}
\affiliation{Centro de Astronomia e Astrof\'{\i}sica da Universidade de Lisboa, Campo
Grande, Edific\'{\i}o C8, 1749-016 Lisboa, Portugal}
\author{M. K. Mak}
\email{mkmak@vtc.edu.hk}
\affiliation{Department of Computing and Information Management, Hong Kong Institute of
Vocational Education, Chai Wan, Hong Kong, P. R. China}
\date{\today}

\begin{abstract}
The mechanism of the initial inflationary scenario of the universe and of
its late-time acceleration can be described by assuming the existence of
some gravitationally coupled scalar fields $\phi $, with the inflaton field
generating inflation and the quintessence field being responsible for the
late accelerated expansion. Various inflationary and late-time accelerated
scenarios are distinguished by the choice of an effective self-interaction
potential $V(\phi )$, which simulates a temporarily non-vanishing
cosmological term. In this work, we present a new formalism for the analysis
of scalar fields in flat isotropic and homogeneous cosmological models. The
basic evolution equation of the models can be reduced to a first order
non-linear differential equation. Approximate solutions of this equation can
be constructed in the limiting cases of the scalar field kinetic energy and
potential energy dominance, respectively, as well as in the intermediate
regime. Moreover, we present several new accelerating and decelerating exact
cosmological solutions, based on the exact integration of the basic
evolution equation for scalar field cosmologies. More specifically, exact
solutions are obtained for exponential, generalized cosine hyperbolic, and
power law potentials, respectively. Cosmological models with power law
scalar field potentials are also analyzed in detail.
\end{abstract}

\maketitle


\section{Introduction}

Scalar fields are assumed to play a fundamental role in cosmology, where one
of the first major mechanisms for which scalar fields are thought to be
responsible is the inflationary scenario \cite{1, 2}. Although originally
inflationary models were proposed in cosmology to provide solutions to the
issues of the singularity, flat space, horizon, homogeneity problems and
absence of magnetic monopoles, as well as to the problem of large numbers of
particles \cite{Li90, Li98}, by far the most useful property of inflation is
that it generates both density perturbations and gravitational waves. These
can be measured in a variety of different ways including the analysis of
microwave background anisotropies, velocity flows in the Universe,
clustering of galaxies and the abundance of gravitationally bound objects of
various types \cite{Li98}. The possibility that a canonical scalar field
with a potential, dubbed \textit{quintessence}, may be responsible for the
late-time cosmic acceleration, was also explored \cite{caldwell}. Contrary
to the cosmological constant, the quintessence equation of state changes
dynamically with time \cite{11}. In fact, a plethora of exotic fluids have
been proposed to explain the accelerated expansion of the Universe, which
include amongst many others $k-$essence models, in which the late-time can
be driven by the kinetic energy of the scalar field \cite{kessence}; coupled
models where dark energy interacts with dark matter \cite{coupledDE}; and
unified models of dark matter and dark energy \cite{DM_DE}.

In a wide range of inflationary models the underlying dynamics is that of a
single scalar field, with the inflaton rolling in some underlying potential %
\cite{1,2,Li90, Li98}. In order to study the inflationary dynamics, the
usual strategy is an expansion in the deviation from the scale invariance,
formally expressed as the slow-roll approximation, which arises in two
separate contexts. The first is in simplifying the classical inflationary
dynamics of expansion and the lowest order approximation ignores the
contribution of the kinetic energy of the inflation to the expansion rate.
The second is in the calculation of the perturbation spectra, where the
standard expressions are valid to lower order in the slow roll approximation %
\cite{Co94}. Exact inflationary solutions have also been found for a large
number of inflationary potentials, and the respective potentials allowing a
graceful exit have been classified \cite{Mi}. In fact, many quintessential
potentials have been proposed in the literature, which may be crudely
classified as ``freezing'' models and ``thawing'' models \cite%
{Caldwell:2005tm}. Note that in the former class \cite{freezing}, the field
rolls along the potential in the past, and the movement gradually slowing
down as the system enters the phase of cosmic acceleration. In the latter,
``thawing'' models, the scalar field, possessing a mass of $m_{\phi }$, has
been frozen by the Hubble friction term $H\dot{\phi}$ until recently, and
eventually starts to evolve as $H$ drops below $m_{\phi }$ \cite{thawing}.
Another interesting model involves a double exponential potential, which
requires that the potential becomes shallow, or has a minimum in order to
slow the movement of the scalar field \cite{Barreiro:1999zs}; the latter
behavior has also been exhibited by more general potentials \cite%
{general_pot}.

More recently, the released Planck data of the 2.7 full sky survey \cite%
{P1,P2} have shown a number of novel and unexpected features, whose
explanation will certainly require a deep change in our standard
understanding of the Universe. These recent observations have measured the
Cosmic Microwave Background to an unprecedented precision. Even though
generally the Planck data confirm the foundations of the $\Lambda $CDM
model, the observational data show some tension between the fundamental
principle of the model and observations. For example, the Planck data
combined with the WMAP polarization data show that the index of the power
spectrum is given by $n_s=0.9603\pm0.0073 $ \cite{P1}, which rules out the
exact scale-invariance ($n_s = 1$) at more than $5\sigma $ level. Hence
Planck data ``severely limits the extensions of the simplest paradigm'' \cite%
{P1}. On the other hand Planck data do not require the consideration of
inflationary models beyond the simplest canonical single field scenarios %
\cite{3}. More specifically, a chaotic inflationary model, based on a
quartic potential, is highly disfavored by the observations. The
inflationary model based on a quadratic potential is marginally consistent
with the observation at 2$\sigma $ level, and models with a linear or
fractional power potential lie outside the 1$\sigma $, but within the 2$%
\sigma $ allowed region \cite{P1, 4}.

The observations of high redshift supernovae, and the WMAP/Planck data,
showing that the location of the first acoustic peak in the power spectrum
of the microwave background radiation is consistent with the inflationary
prediction $\Omega =1$, have provided compelling evidence for a net equation
of state of the cosmic fluid lying in the range $-1\leq w=p/\rho <-1/3$ \cite%
{acc}. To explain these observations, two dark components are invoked:
pressureless cold dark matter (CDM), and dark energy (DE) with negative
pressure. CDM contributes $\Omega _{m}\sim 0.3$ \cite{P2}, and is mainly
motivated by the theoretical interpretation of the galactic rotation curves
and large scale structure formation. DE is assumed to provide $\Omega
_{DE}\sim 0.7$, and is responsible for the acceleration of the distant type
Ia supernovae \cite{acc}. There are a huge number of proposed candidates for
DE (see, for instance, \cite{PeRa03, Pa03}). One possibility is cosmologies
based on a mixture of cold dark matter and quintessence, a slowly-varying,
spatially inhomogeneous component \cite{8}. An example of implementation of
the idea of quintessence is the suggestion that it is the energy associated
with a scalar field $Q$, with a self-interaction potential $V(Q)$. If the
potential energy density is greater than the kinetic one, then the pressure $%
p=\dot{Q}^{2}/2-V(Q)$ associated to the $Q$-field is negative.
Quintessential cosmological models have been intensively investigated in the
physical literature (for a recent review see \cite{Tsu}). The interaction
between dark energy and dark matter in the framework of irreversible
thermodynamics of open systems with matter creation/annihilation has also
been recently explored \cite{Harko:2012za}, where dark energy and dark
matter are considered as an interacting two component (scalar field and
``ordinary'' dark matter) cosmological fluid in a homogeneous spatially flat
and isotropic Friedmann-Robertson-Walker (FRW) Universe. The possibility of
cosmological anisotropy from non-comoving dark matter and dark energy have
also been proposed \cite{Harko:2013wsa}.

Models with nonstandard scalar fields, such as phantom scalar fields and
Galileons, which can have bounce solutions and dark energy solutions with $%
w<-1$ have also been extensively investigated in the literature. In the
Galileon theory one imposes an internal Galilean invariance, under which the
gradient of the relativistic scalar field $\pi $, with peculiar derivative self-interactions, and universally coupled to matter,
 is shifted by a constant term \cite{gal1}. The Galilean symmetry constrains the structure of the Lagrangian of the scalar field
 so that in four dimensions  only five terms
that can yield sizable non-linearities without introducing ghosts do exist. Different
extensions of the Galileon models were considered in \cite{gal2}. In \cite%
{gal3} a new class of inflationary models was proposed, in which the
standard model Higgs boson can act as an inflaton due to Galileon-like
non-linear derivative interaction. The generated primordial density
perturbation is consistent with the present observational data. Generalized
Galileons as a framework to develop the most general single-field inflation
models, i.e., Generalized G-inflation, were studied in \cite{gal4}. As special cases this
model contains k-inflation, extended inflation, and new Higgs inflation. The background and perturbation evolution in this model were
investigated, and the most general quadratic actions for tensor and scalar
cosmological perturbations was obtained. The stability criteria and the
power spectra of primordial fluctuations were also presented. For a recent
review of scalar field theories with second-derivative Lagrangians, whose
field equations are second order see \cite{Rub}. Some of these theories
admit solutions violating the null energy condition and have no obvious
pathologies.

In order to explain the recent acceleration of the Universe, in which $w<-1$%
, scalar fields $\phi $ that are minimally coupled to gravity with a
negative kinetic energy, and which are known as ``phantom fields'', have
been introduced in \cite{phan1}. The energy density and pressure of a
phantom scalar field are given by $\rho _{\phi}=-\dot{\phi}^2/2+V\left(\phi
\right)$ and $p _{\phi}=-\dot{\phi}^2/2-V\left(\phi \right)$, respectively.
The properties of phantom cosmological models have been investigated in \cite%
{phan2}. The phenomenon of the phantom divide line crossing in the scalar
field models with cusped potentials was considered in \cite{phan3}.
Cosmological observations show that at some moment in the past the value of the parameter $w$ of the dark energy equation of state
may have crossed the value $w= -1$,
corresponding to the cosmological constant $\Lambda $. This phenomenon is called
 phantom divide line crossing \cite{phan4}.  Non-phantom dark energy is described by a minimally coupled
scalar field, having a kinetic term with the
positive sign. Therefore in order to describe
the phantom divide line crossing  it seems natural to use two scalar fields, a
phantom field with the negative kinetic term, and a standard one \cite{phan3}. Another possible way of
explaining the phantom divide line crossing is to use a scalar field
nonminimally coupled to gravity \cite{phan3}.

The mathematical properties of the Friedmann-Robertson-Walker (FRW)
cosmological models with a scalar field as matter source have also been
intensively investigated. In \cite{Mus, Sal} a simple way of reducing the
system of the gravitational field equations to one first order equation was
proposed, namely, to the Hamilton-Jacobi-like equation for the Hubble
parameter $H$considered as a function of the scalar field $\phi $, $%
3H^{2}\left( \phi \right) =V(\phi )+2(dH/d\phi )^{2}$. The gravitational
collapse and the dynamical properties of scalar field models were considered
in \cite{Giambi}. The solution of the field equations for a cosine
hyperbolic type scalar field potential for the case of an equation of state
equivalent to the nonrelativistic matter plus a cosmological term was
derived in \cite{Kis}. The relation between the inflationary potential and
the spectra of density (scalar) perturbations and gravitational waves
(tensor perturbations) produced during inflation, and the possibility of
reconstructing the inflaton potential from observations, was considered in %
\cite{Lid0}. If inflation passes a consistency test, one can use
observational information to constrain the inflationary potential. The key
point in the reconstruction procedure is that the Hubble parameter is
considered as a function of the scalar field, and this allows to reconstruct
the scalar field potential, and determine the dynamics of the field itself,
without a priori knowing the Hubble parameter as a function of time or of
the scale factor \cite{Kam1,Kam2}. General solutions for flat Friedmann
universes filled with a scalar field in induced gravity models and models
including the Hilbert-Einstein curvature term plus a scalar field
conformally coupled to gravity were also derived in \cite{Kam3}. The
corresponding models are connected with minimally coupled solutions through
the combination of a conformal transformation and a transformation of the
scalar field. The explicit forms of the self-interaction potentials for six
exactly solvable models was also obtained. In \cite{Nunes:2000yc}, a
phase-plane analysis was performed of the complete dynamical system
corresponding to a flat FRW cosmological models with a perfect fluid and a
self-interacting scalar field and it was shown that every positive and
monotonous potential which is asymptotically exponential yields a scaling
solution as a global attractor. The dynamics of models of warm inflation
with general dissipative effects, was also extensively analyzed \cite%
{Mimoso:2005bv}, and a mechanism that generates the exact solutions of
scalar field cosmologies in a unified way was also investigated.

The connections between the Korteweg-de Vries equation and inflationary
cosmological models were explored in \cite{Lid1}. The relation between the
non-linear Schr\"{o}dinger equation and the cosmological Friedmann equations
for a spatially flat and isotropic Universe in the presence of a
self-interacting scalar field has been considered in \cite{Lid2}. A
Hamiltonian formalism for the study of scalar fields coupled to gravity in a
cosmological background was developed in \cite{Ber}. A number of integrable
one--scalar spatially flat cosmologies, which play a natural role in the
inflationary scenarios, were studied in \cite{Fre1}. Systems with potentials
involving combinations of exponential functions, and similar non--integrable
cases were also studied in detail. It was shown that the scalar field
emerges from the initial singularity while climbing up sufficiently steep
exponential potentials (``climbing phenomenon''), and that it inevitably
collapses in a Big Crunch, whenever the scalar field tries to settle at the
negative extremals of the potential. The question whether the integrable one
scalar-field cosmologies can be embedded as consistent one-field truncations
into Extended Gauged Supergravity or in $N=1$supergravity gauged by a
superpotential without the use of $D-$terms was considered in \cite{Fre2}.

Therefore, the theoretical investigation of scalar field models is an
essential task in cosmology. It is the purpose of the present paper to
consider a systematic analysis of scalar field cosmologies, and to derive a
basic evolution equation describing flat, isotropic and homogeneous scalar
field cosmological models. The evolution equation is a first order, strongly
non-linear differential equation, which, however, allows the possibility of
considering analytical solutions in both the asymptotic limits of scalar
field kinetic or potential energy dominance and in the intermediate domain,
respectively. Moreover, a large number of exact solutions can also be
obtained. The cases of the exponential, hyperbolic cosine, and power law
potentials are explicitly considered.

The present paper is organized as follows. The basic evolution equation for
scalar field cosmologies with an arbitrary self-interaction potential is
derived in Section \ref{sect2}. Several classes of exact scalar field
solutions are considered in Sections \ref{sect3} and \ref{secn}. The general
formalism is used in Section \ref{sect5} to obtain some approximate
solutions of the gravitational field equations. In Section \ref{sect6} we
consider in detail the case of the simple power law potential. We discuss
and conclude our results in Section \ref{sect7}.

\section{Scalar field cosmologies with arbitrary self-interaction potential}

\label{sect2}

Let us consider a rather general class of scalar field models, minimally
coupled to the gravitational field, for which the Lagrangian density in the
Einstein frame reads
\begin{equation}
L=\frac{1}{2\kappa}\sqrt{\left| g\right| }\left\{ R+\kappa \left[ g^{\mu \nu
}\left( \partial _{\mu }\phi \right) \left( \partial _{\nu }\phi \right)
-2V\left( \phi \right) \right] \right\} \,,
\end{equation}
where $R$ is the curvature scalar, $\phi $ is the scalar field, $V\left(
\phi \right) $ is the self-interaction potential and $\kappa =8\pi G/c^{4}$
is the gravitational coupling constant, respectively. In the following, we
use natural units with $c=8\pi G=\hbar =1$, and we adopt as our signature
for the metric $\left( +1,-1,-1,-1\right) $, as is common in particle
physics.

For a flat FRW scalar field dominated Universe with the line element
\begin{equation}
ds^{2}=dt^{2}-a^{2}(t)\left( dx^{2}+dy^{2}+dz^{2}\right) ,
\end{equation}%
where $a$ is the scale factor, the evolution of a cosmological model is
determined by the system of the field equations
\begin{eqnarray}
3H^{2} &=&\rho _{\phi }=\frac{\dot{\phi}^{2}}{2}+V\left( \phi \right) ,
\label{H} \\
2\dot{H}+3H^{2} &=&-p_{\phi }=-\frac{\dot{\phi}^{2}}{2}+V\left( \phi \right),
\label{H1}
\end{eqnarray}%
and the evolution equation for the scalar field
\begin{equation}
\ddot{\phi}+3H\dot{\phi}+V^{\prime }\left( \phi \right) =0,  \label{phi}
\end{equation}%
where $H=\dot{a}/a>0$ is the Hubble expansion rate function,  the overdot
denotes the derivative with respect to the time-coordinate $t$, while the prime
denotes the derivative with respect to the scalar field $\phi $,
respectively. In the following we will restrict our study to expansionary cosmological models, which satisfy the condition that the scale factor is a monotonically increasing function of time. For expanding cosmological models the condition $H>0$ is always satisfied. Cosmological models with $H<0$ correspond to collapsing scalar field configurations, in which the scale factor is a monotonically decreasing function of time.

By adding Eqs.~(\ref{H}) and (\ref{H1}), we obtain the Riccati
type equation satisfied by $H$, of the form
\begin{equation}
\dot{H}=V-3H^{2}.  \label{a3}
\end{equation}

By substituting the Hubble function from Eq.~(\ref{H}) into Eq.~(\ref{phi}),
we obtain the basic equation describing the scalar field evolution as
\begin{equation}
\ddot{\phi}+\sqrt{3}\sqrt{\frac{\dot{\phi}^{2}}{2}+V\left( \phi \right) }\;
\dot{\phi}+\frac{dV}{d\phi }=0.  \label{in}
\end{equation}

Now, in order to deduce a basic equation describing the dynamics of the
scalar fields in the flat FRW Universe, which will be useful throughout this
work, we consider several transformations. First, by defining a new function
$f\left( \phi \right)$ so that $\dot{\phi}=\sqrt{f\left( \phi \right) }$,
and changing the independent variable from $t$ to $\phi $, Eq. (\ref{in})
becomes
\begin{equation}
\frac{1}{2}\frac{df\left( \phi \right) }{d\phi }+\sqrt{3}\sqrt{\frac{f\left(
\phi \right) }{2}+V\left( \phi \right) }\sqrt{f\left( \phi \right) }%
+V^{\prime }\left( \phi \right) =0,  \label{in1}
\end{equation}%
which may be reorganized into the following form:
\begin{equation}
\frac{\frac{1}{2}\frac{df\left( \phi \right) }{d\phi }+V^{\prime }\left(
\phi \right) }{2\sqrt{\frac{f\left( \phi \right) }{2}+V\left( \phi \right) }}%
+\frac{\sqrt{3}}{2}\sqrt{f\left( \phi \right) }=0.  \label{in2}
\end{equation}

Next, by introducing a new function $F(\phi )=\sqrt{f\left( \phi \right)
/2+V\left( \phi \right) }$, so that $f(\phi )=2\left[ F^{2}(\phi )-V(\phi )%
\right] $, Eq.~(\ref{in2}) takes the form
\begin{equation}
\frac{dF(\phi)}{d\phi }+\sqrt{\frac{3}{2}}\sqrt{V(\phi)}\sqrt{\left[ \frac{%
F(\phi)}{\sqrt{V(\phi)}}\right] ^{2}-1}=0.  \label{in3}
\end{equation}
Thus, we introduce now the function $u(\phi)$ defined as $F(\phi )=u(\phi)%
\sqrt{V(\phi)}$, which transforms Eq.~(\ref{in3}) to
\begin{equation}
\frac{1}{\sqrt{u^{2}-1}}\frac{du}{d\phi }+ \frac{1}{2V}\frac{dV}{d\phi }%
\frac{u}{\sqrt{u^{2}-1}}+\sqrt{\frac{3}{2}}=0.
\end{equation}

With the help of the transformation $u(\phi )=\cosh G(\phi )$, we obtain the
basic equation describing the dynamics of the scalar fields in the flat FRW
Universe as
\begin{equation}
\frac{dG}{d\phi }+\frac{1}{2V}\frac{dV}{d\phi }\coth G+\sqrt{\frac{3}{2}}=0.
\label{fin}
\end{equation}%
For $f$ we obtain $f(\phi )=2V(\phi )\sinh ^{2}G(\phi )$, leading to $\dot{%
\phi}=\sqrt{2V\left( \phi \right) }\sinh G\left( \phi \right) $. As a
function of time $G$ satisfies the equation
\begin{equation}
\frac{dG}{dt}=-\sqrt{2V\left( \phi \right) }\sinh G\left[ \sqrt{\frac{3}{2}}+%
\frac{1}{2V\left( \phi \right) }\frac{dV}{d\phi }\coth G\right] .
\label{time}
\end{equation}

Note that the function $G$ can be obtained from the scalar field with the
use of the equation
\begin{equation}  \label{G}
G(\phi)=\mathrm{arccosh} \sqrt{1+\frac{\dot{\phi}^{2}}{2V(\phi)}}.
\end{equation}
Furthermore, as a function of the scalar field, the scale factor $a$ is
given by the equation
\begin{equation}  \label{a}
\frac{1}{a(\phi)}\frac{da(\phi)}{d\phi }=\frac{1}{\sqrt{6}}\coth G\left(
\phi \right) .
\end{equation}
or alternatively, as a function of $G$, the latter scale factor can be
obtained from
\begin{equation}
\frac{1}{a}\frac{da}{dG}=-\frac{1}{\sqrt{6}}\frac{\coth G}{\sqrt{\frac{3}{2}}%
+\frac{1}{2V}\frac{dV}{d\phi }\coth G}.
\end{equation}

An important observational quantity, the deceleration parameter $q$, can
also be expressed in the form
\begin{equation}
q(\phi )=\frac{d}{dt}\left( \frac{1}{H}\right) -1=\sqrt{6\left[ F^{2}(\phi
)-V(\phi )\right] }\frac{d}{d\phi }\left[ F(\phi )\right] ^{-1}-1.
\end{equation}%
As a function of the potential $V$ and of $G$, the deceleration parameter is
given by
\begin{equation}
q(\phi )=\sqrt{6V(\phi )}\sinh G(\phi )\frac{d}{d\phi }\left[ \frac{1}{\sqrt{%
V(\phi )}\cosh G(\phi )}\right] -1.  \label{aa}
\end{equation}%
By inserting Eq.~(\ref{fin}) into Eq.~(\ref{aa}), the latter yields the
result
\begin{equation}
q(\phi )=3\tanh ^{2}G(\phi )-1,  \label{bb}
\end{equation}%
and by substituting Eq.~(\ref{G}) into Eq.~(\ref{bb}), yields the
deceleration parameter in the following useful form:
\begin{equation}
q(\phi )=2-3\left( 1+\frac{\dot{\phi}^{2}}{2V(\phi )}\right) ^{-1},
\label{q1}
\end{equation}%
respectively. If the potential energy dominates the kinetic energy of the
scalar field, $V(\phi )\gg \dot{\phi}^{2}/2$, from Eq.~(\ref{q1}) it follows
that $q\rightarrow -1$, and it is important to note that this property is
independent of the form of the scalar field self-interaction potential $%
V(\phi )$.

By using the new variable $f(\phi)$ and $F(\phi)$ from Eqs.~(\ref{H}) and (\ref{H1}) we obtain
\be\label{21}
\dot {H}=-\frac{\dot{\phi }^2}{2}=-\frac{f(\phi)}{2}
\ee
and
\be\label{22}
3H^2=F^2(\phi),
\ee
respectively. Eqs.~(\ref{21}) and (\ref{22}) show that the functions $f(\phi)$ and $F(\phi)$ are related to the Hubble function, and its time derivative. A similar approach, in which the Hubble function is assumed to be a function of the scalar field $\phi $, was considered in \cite{Sal}.

\section{Exact scalar field models}

\label{sect3}

As mentioned in the Introduction, scalar fields are considered to play a
central role in current models of the early Universe. The self-interaction
potential energy density of such a field is undiluted by the expansion of
the Universe, and hence can behave as an effective cosmological constant,
driving a period of inflation, or of a late-time acceleration. The evolution
of the Universe is strongly dependent upon the specific form of the scalar
field potential $V(\phi)$. A common form for the self-interaction potential
is the exponential type potential. Note that Eq.~(\ref{fin}) can be
integrated immediately in the case of potentials satisfying the condition $%
V^{\prime }/V=\mathrm{constant}$. Therefore, for this class of potentials
the general solution of the gravitational field equations can be obtained in
an exact analytical form. Other classes of exact solutions can be
constructed by assuming that $V^{\prime }/V$ is some function of $G$, i.e., $%
V^{\prime }/V=f\left(G\right)$. For a large number of choices of the
function $f(G)$, the first order evolution equation, given by Eq.~(\ref{fin}%
), can be solved exactly, and the solution corresponding to a given
potential can be obtained in an exact form. In the following, we consider
some exact analytical classes of scalar field cosmologies.

\subsection{The exponential potential scalar field}

If $V^{\prime }/V$ is a constant, that is, $V^{\prime }/V=\sqrt{6}\alpha
_{0}= $constant, the scalar field self-interaction potential is of the
exponential form,
\begin{equation}
V=V_{0}\exp \left( \sqrt{6}\alpha _{0}\phi \right) ,  \label{pp}
\end{equation}%
where $V_{0}$ is an arbitrary constant of integration. The cosmological
behavior of the Universe filled with a scalar field, with a Liouville-type
exponential potential, has been extensively investigated in the physical
literature for both homogeneous and inhomogeneous scalar fields \cite%
{exp1,K1,K2,Cui}. In particular, an exponential potential arises in
four-dimensional effective Kaluza-Klein type theories from compactification
of the higher-dimensional supergravity or superstring theories \cite%
{CaMaPeFr85}. In string or Kaluza-Klein theories the moduli fields
associated with the geometry of the extra-dimensions may have effective
exponential potentials due to the curvature of the internal spaces or to the
interaction of the moduli with form fields on the internal spaces.
Exponential potentials can also arise due to non-perturbative effects such
as gaugino condensation \cite{CaCaMu93}. The integrability of the
gravitational field equations for exponential type scalar potentials was
considered in \cite{Sch,Sch1,Sch2,pow1}.

Taking into account Eq.~(\ref{pp}), then Eq. (\ref{fin}) takes the form
\begin{equation}
\frac{dG}{d\phi }+\sqrt{\frac{3}{2}}\left( \alpha _{0}\coth G+1\right) =0.
\label{dd}
\end{equation}%
We analyze below several cases of interest.

\subsubsection{The case $\protect\alpha _0\neq \pm 1$}

For $\alpha _0\neq \pm 1$, Eq.~(\ref{dd}) gives immediately
\begin{equation}
\sqrt{\frac{3}{2}}\left[ \phi (G)-\phi _{0}\right] =\frac{G-\alpha _{0}\ln
\left| \sinh G+\alpha _{0}\cosh G\right| }{\alpha _{0}^{2}-1}\,,
\label{expphi}
\end{equation}%
where $\phi _{0}$ is an arbitrary constant of integration. The time
dependence of the physical parameters can be obtained from Eq.~(\ref{time})
as
\begin{equation}
t(G)-t_{0}=-\frac{1}{\sqrt{3V_{0}}}\int {\frac{dG}{e^{\sqrt{3/2}\alpha
_{0}\phi }\left( \sinh G+\alpha _{0}\cosh G\right) }}.  \label{exp1b}
\end{equation}

With the use of Eq.~(\ref{expphi}), we obtain the following integral
representation for the time $t$,
\bea  \label{exp2}
&&t(G)-t_{0}=-\frac{e^{-\sqrt{3/2}\alpha _{0}\phi _{0}}}{\sqrt{3V_{0}}}\times\nonumber\\
&&\int {%
e^{\left[ \alpha _{0}/\left( 1-\alpha _{0}^{2}\right) \right] G}\left( \sinh
G+\alpha _{0}\cosh G\right) ^{1/\left( \alpha _{0}^{2}-1\right) }dG}, \nonumber\\
\eea
where $t_{0}$ is an arbitrary constant of integration. Thus, Eqs.~(\ref%
{exp1b}) and (\ref{exp2}) give a parametric representation of the time
evolution of the scalar field $\phi $ with an exponential type
self-interaction potential, with $G$ taken as a parameter. For the scale
factor, we obtain
\begin{equation}
a(G)=a_{0}e^{\left[ \alpha _{0}/3\left( 1-\alpha _{0}^{2}\right) \right]
G}\left( \sinh G+\alpha _{0}\cosh G\right) ^{1/3\left( \alpha
_{0}^{2}-1\right) },
\end{equation}
where $a_{0}$ is an arbitrary constant of integration, while the
deceleration parameter is given by
\begin{equation}
q(G)=3\tanh ^{2}G-1.  \label{qexp}
\end{equation}

The time integral given by Eq.~(\ref{exp2}) can be obtained in an exact form
for some particular values of $\alpha _{0}$. Thus, if $\alpha _{0}=\pm \sqrt{%
2}$, we have
\begin{equation}
t_{\pm }(G)-t_{0}^{\pm }=-\frac{e^{\mp \left( \sqrt{2}G+\sqrt{3}\phi
_{0}\right) }}{\sqrt{3V_{0}}}\left[ 3\cosh G\pm 2\sqrt{2}\sinh G\right] .
\end{equation}

For $\alpha _{0}=\pm \sqrt{3/2}$ we obtain
\bea
t_{\pm }(G)-t_{0}^{\pm }=\pm \frac{1}{24\sqrt{V_{0}}}\mathbf{e}^{\mp \left(
\sqrt{6}G+\frac{3\phi _{0}}{2}\right) } \times
   \nonumber\\
 \times \left[ \sqrt{2}+ 27\sqrt{2}\cosh
(2G)\pm 22\sqrt{3}\sinh (2G)\right],
\eea
while for $\alpha _{0}=\pm 2/\sqrt{3}$ we find
\begin{eqnarray}
t_{\pm }(G)-t_{0}^{\pm } =\frac{1}{396\sqrt{3V_{0}}}e^{\mp \left( 2\sqrt{3}%
G+\sqrt{2}\phi _{0}\right) }\Big\{ 45\cosh G\nonumber\\
+1067\cosh (3G)\pm 8\sqrt{3}%
\left[ 3\sinh G+77\sinh (3G)\right] \Big\} .
\end{eqnarray}

Therefore, in these cases the exact solution of the gravitational field
equations in the presence of a scalar field with exponential potential can
be obtained in an exact parametric form, and there is no need to resort to
numerical integration.

A particular solution of the field equations corresponds to the case $G(\phi
)=G_{0}=\mathrm{constant}$. In this case Eq.~(\ref{dd}) is identically
satisfied, with $G_{0}$ given by
\begin{equation}
G_{0}=\mathrm{arccoth}\left( -\frac{1}{\alpha _{0}}\right) =\frac{1}{2}\ln
\left| \frac{1-\alpha _{0}}{1+\alpha _{0}}\right| , 0<|\alpha_0|<1.
\end{equation}

From Eq.~(\ref{a}) it follows that the scale factor can be obtained as a
function of the scalar field as
\begin{equation}
a=a_{0}e^{-\phi /\sqrt{6}\alpha _{0}},  \label{41}
\end{equation}%
while the time variation of the scalar field is determined from Eq.~(\ref{G}%
) by the equation
\begin{equation}
\dot{\phi}=\pm \sqrt{2V_{0}}\sinh \left( G_{0}\right) e^{\sqrt{3/2}\alpha
_{0}\phi },
\end{equation}%
with the general solution given by
\begin{equation}
e^{-\sqrt{3/2}\alpha _{0}\phi }=\pm \sqrt{3V_{0}}\alpha _{0}\sinh \left(
G_{0}\right) \left( t_{0}-t\right) ,
\end{equation}%
where $t_{0}$ is an arbitrary integration constant.
With the help of Eq.~(\ref{41}) we obtain the scale factor in the form
\begin{equation}
a(t)=a_{0}\left[ \pm \sqrt{3V_{0}}\alpha _{0}\sinh \left( G_{0}\right)
\left( t_0-t\right)%
\right] ^{\frac{1}{3\alpha _{0}^{2}}}.
\end{equation}

The time variation of the scalar field potential is given by
\bea\label{potsimpl}
V(t)&=&\frac{V_{0}}{3V_{0}\alpha _{0}^{2}\sinh ^{2}\left( G_{0}\right) }\frac{1%
}{\left( t-t_{0}\right) ^{2}}\nonumber\\
&=&\left( \frac{1-\alpha _{0}^{2}}{3\alpha
_{0}^{4}}\right) \frac{1}{\left( t-t_{0}\right) ^{2}}=\frac{V_{0}}{\left( t-t_{0}\right) ^{2}},
\eea
with the constants $V_{0}$, $\alpha _{0}$ and $%
G_{0}$ satisfying the consistency condition %
\begin{equation}
3V_{0}\alpha _{0}^{2}\sinh ^{2}\left(G_{0}\right)=1,
\end{equation}
which follows immediately from the comparison of the second and the last term in Eq.~(\ref{potsimpl}).

Simple power law solutions for cosmological models with scalar fields with
exponential potentials have been obtained, and studied, in \cite{pow1}.

\begin{figure*}
\centering
\includegraphics[width=8.5cm, height=5.5cm]{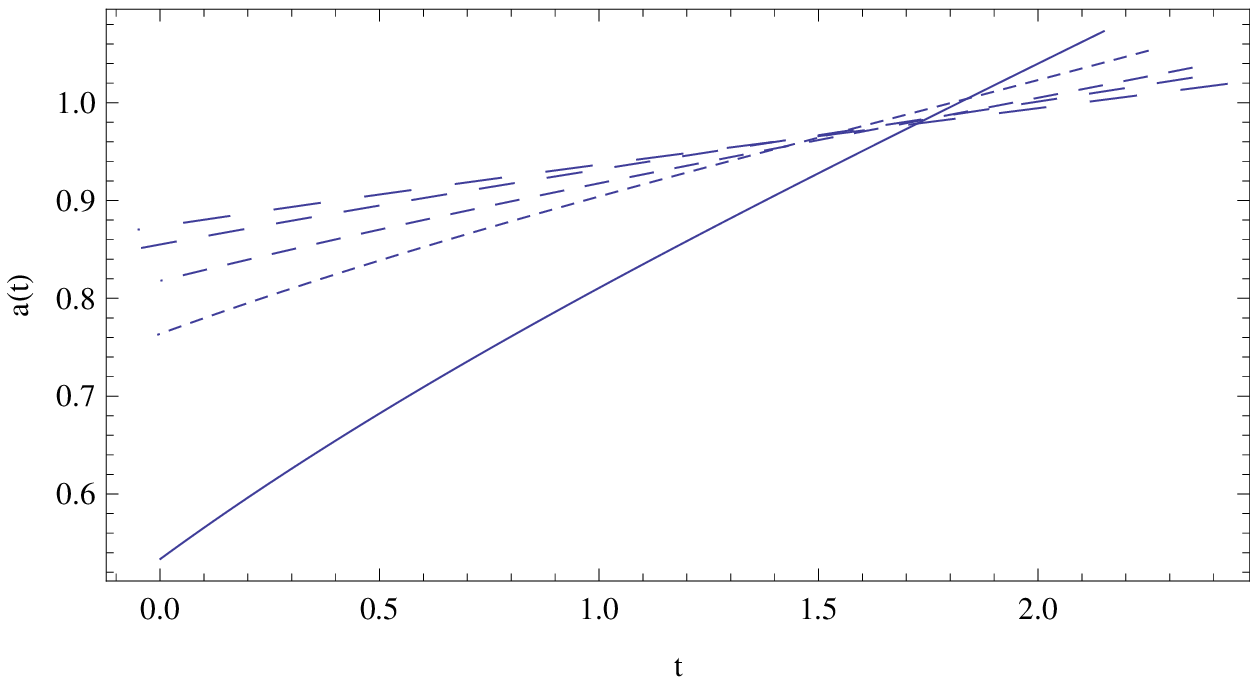}
\includegraphics[width=8.5cm, height=5.5cm]{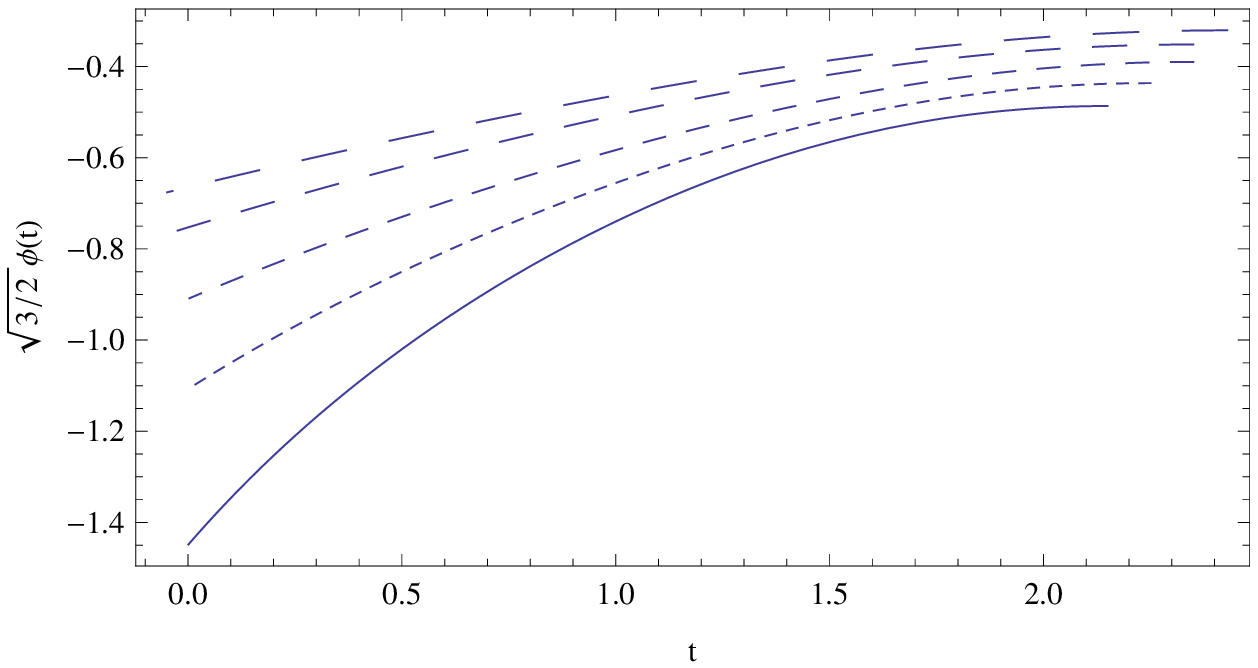}
\caption{Depicted is the time variation of the scale factor (in arbitrary units), in the left plot, and the time variation of the cosmological  scalar field, in the right plot,  with an exponential  potential for different values of $\alpha _0$: $\alpha _0=1.5 $ (solid curve), $\alpha _0=2.5$ (dotted curve), $\alpha _0=3.5$ (short dashed curve), $\alpha _0=4.5$ (dashed curve), and $\alpha _0=5.5$ (long dashed curve), respectively. The arbitrary integration constants $\phi _0$ and $V_0$ have been normalized so that $\exp \left(-\sqrt{3/2}\alpha _0\phi _0\right)=\sqrt{3V_0}$. }\label{Vexp1}
\end{figure*}

\begin{figure*}
\centering
\includegraphics[width=8.5cm, height=5.5cm]{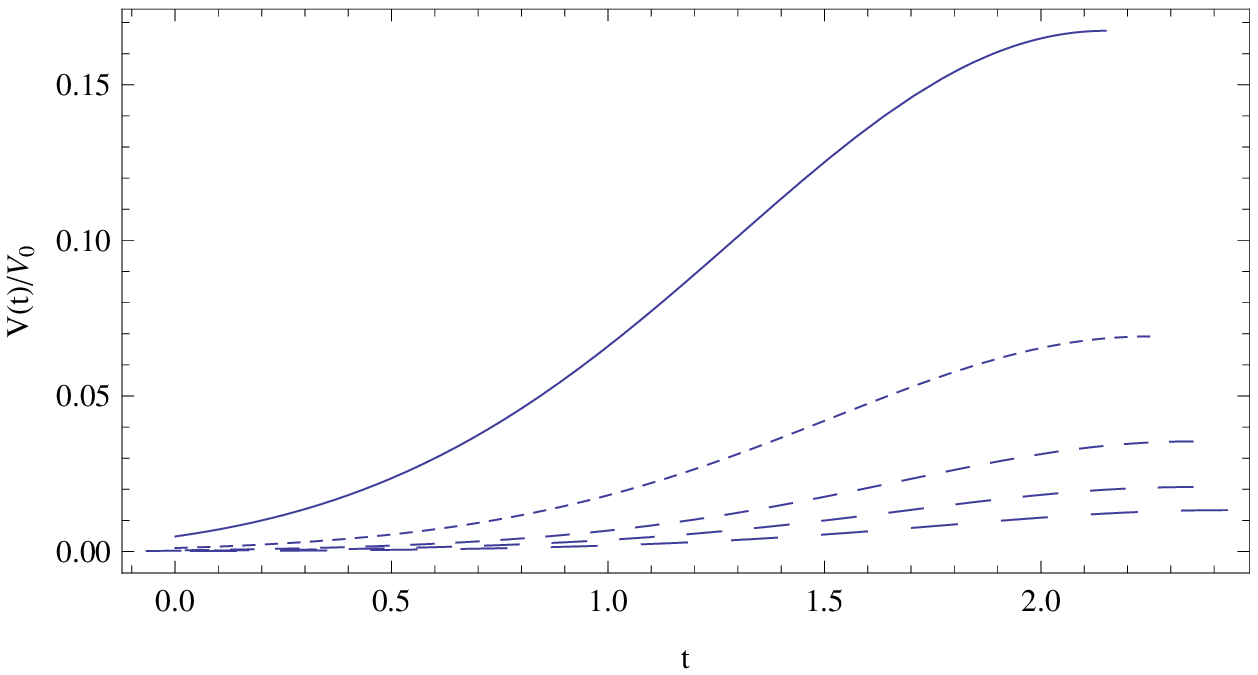}
\includegraphics[width=8.5cm, height=5.5cm]{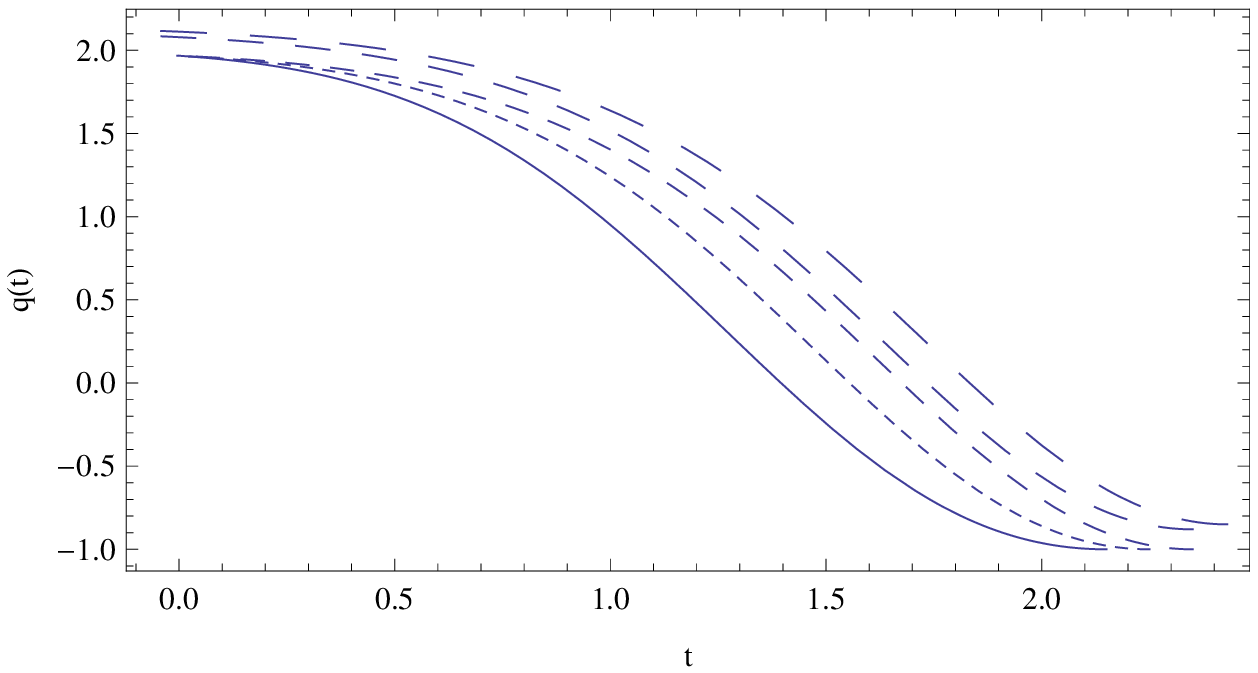}
\caption{Plots of the time variation of the exponential  scalar field potential, depicted in the left figure, and the
time variation of the deceleration parameter of the Universe filled with an exponential potential scalar field, depicted in the right figure,  for different values of $\alpha _0$: $\alpha _0=1.5 $ (solid curve), $\alpha _0=2.5$ (dotted curve), $\alpha _0=3.5$ (short dashed curve), $\alpha _0=4.5$ (dashed curve), and $\alpha _0=5.5$ (long dashed curve), respectively. The arbitrary integration constants $\phi _0$ and $V_0$ have been normalized so that $\exp \left(-\sqrt{3/2}\alpha _0\phi _0\right)=\sqrt{3V_0}$.}
\label{exp4}
\end{figure*}

The time variations of the scale factor, scalar field, scalar field
potential, and deceleration parameter for the exponential potential scalar
field filled Universe are represented, for different values of $\alpha _0$,
in Figs.~\ref{Vexp1}-\ref{exp4}. For the considered range of the parameter $%
\alpha _0$, the scale factor is a monotonically increasing function of time,
and therefore the solution represents an expanding Universe. The Universe
starts its evolution from a decelerating phase, with $q>0$, but after a
finite interval it enters in an accelerated era, with $q<0$. In the large
time limit $t\rightarrow \infty $, as can be seen immediately from Eq.~(\ref%
{expphi}), for the chosen values of the parameters, the scalar field $\phi $
is an increasing function of time, becoming a constant in the large time
limit, as well as the scalar field potential. In the limit of large times
the Universe enters a de Sitter type accelerated phase, with $q=-1$.

\subsubsection{The case $\alpha _0=\pm 1$}

In the specific case of $\alpha _{0}=\pm 1$, from Eq.~(\ref{dd}) we obtain
the following dependence of the scalar field on $G$,
\begin{equation}
\sqrt{24}\left[ \phi (G)-\phi _{0}^{+}\right] =-e^{-2G}-2G, \qquad \alpha
_{0}=+1,
\end{equation}%
and
\begin{equation}
\sqrt{24}\left[ \phi (G)-\phi _{0}^{-}\right] =\ln \left| \frac{\coth G-1}{%
\coth G+1}\right| +e^{2G}-1, \quad \alpha _{0}=-1,
\end{equation}%
respectively, where $\phi _{0}^{+}$ and $\phi _{0}^{-}$ are arbitrary
constants of integration. For the time dependence of the cosmological model
we obtain the integral representations
\begin{equation}
t(G)-t_{0}^{+}=-\frac{1}{\sqrt{3V_{0}}}\int \frac{e^{-\sqrt{3/2}\phi }}{%
\sinh G+\cosh G}dG, \qquad \alpha _{0}=+1,
\end{equation}%
and
\begin{equation}
t(G)-t_{0}^{-}=-\frac{1}{\sqrt{3V_{0}}}\int \frac{e^{\sqrt{3/2}\phi }}{\sinh
G-\cosh G}dG, \qquad \alpha _{0}=-1,
\end{equation}%
respectively, where $t_{0}^{+}$ and $t_{0}^{-}$ are arbitrary constants of
integration, giving the explicit dependence of the physical time on the
parameter $G$ as
\bea
t(G)-t_{0}^{+}&=&-\frac{e^{-\sqrt{3/2}\phi _{0}^{+}}}{\sqrt{3V_{0}}}%
\int \frac{\exp \left[ (1/4)\left( e^{-2G}+2G\right) \right] }{\sinh G+\cosh
G}dG, \nonumber\\
&& \alpha _{0}=+1,
\eea
and
\bea
&&t(G)-t_{0}^{-}=-\frac{e^{\sqrt{3/2}\phi _{0}^{-}}}{\sqrt{3V_{0}}}\times \nonumber\\
&&\int \frac{%
\left[ (\coth G-1)/(\coth G+1)\right] ^{1/4}\exp \left[ (1/4)\left(
e^{2G}-1\right) \right] }{\sinh G-\cosh G}dG,  \nonumber\\
&&\alpha _{0}=-1,
\eea
respectively. The parametric dependence of the scale factor is given by
\begin{equation}
a(G)=a_{0}^{+}\exp \left[ \frac{1}{12}\left( 2G-e^{-2G}\right) \right] ,
\qquad \alpha _{0}=+1,
\end{equation}
and
\begin{equation}
a(G)=a_{0}^{-}\exp \left[ \frac{1}{12}\left( e^{2G}+2G\right) \right] ,
\qquad \alpha_{0}=-1,
\end{equation}%
respectively, where $a_{0}^{+}$ and $a_{0}^{-}$ are arbitrary constants of
integration. The deceleration parameter is given in parametric form by Eq.~(%
\ref{qexp}).

\subsection{Generalized hyperbolic cosine type scalar field potentials}

As a second example of an exact integrability of the evolution equation,
given by Eq.~(\ref{fin}), we consider the case in which the scalar field
potential can be represented as a function of $G$ in the form
\begin{equation}
\frac{1}{2V}\frac{dV}{d\phi }=\sqrt{\frac{3}{2}}\;\alpha _{1}\,\tanh G,
\end{equation}%
where $\alpha _{1}$ is an arbitrary constant. With this choice, the
evolution equation takes the simple form
\begin{equation}
\frac{dG}{d\phi }=\sqrt{\frac{3}{2}}\left( 1+\alpha _{1}\right) ,
\end{equation}%
with the general solution given by
\begin{equation}
G\left( \phi \right) =\sqrt{\frac{3}{2}}\left( 1+\alpha _{1}\right) \left(
\phi -\phi _{0}\right) ,
\end{equation}%
where $\phi _{0}$ is an arbitrary constant of integration. With the use of
this form of $G$, we immediately obtain the self-interaction potential of
the scalar field as
\begin{equation}
V\left( \phi \right) =V_{0}\cosh ^{\frac{2\alpha _{1}}{1+\alpha _{1}}}\left[
\sqrt{\frac{3}{2}}\left( 1+\alpha _{1}\right) \left( \phi -\phi _{0}\right) %
\right] .  \label{kkk}
\end{equation}

The time dependence of the scalar field can be obtained in a parametric form
as
\begin{widetext}
\be
t-t_{0}=\frac{1}{\sqrt{2V_{0}}} \int \frac{d\phi }{\cosh ^{\frac{\alpha _1}{%
1+\alpha _1}}\left[ \sqrt{\frac{3}{2}}\left( 1+\alpha _1\right) \left( \phi
-\phi _{0}\right) \right] \sinh \left[ \sqrt{\frac{3}{2}}\left( 1+\alpha _1
\right) \left( \phi -\phi _{0}\right) \right] }.
\ee
\end{widetext}
Finally, the scale factor is given by
\begin{equation}
a=a_{0}\sinh ^{\frac{1}{3\left( 1+\alpha _1\right) }}\left[ \sqrt{\frac{3}{2}%
}\left( 1+\alpha _1\right) \left( \phi -\phi _{0}\right) \right] ,
\end{equation}
while the deceleration parameter is expressed as
\begin{equation}
q=3\tanh ^{2}\left[ \sqrt{\frac{3}{2}}\left( 1+\alpha _1\right) \left( \phi
-\phi _{0}\right) \right] -1.
\end{equation}

\begin{centering}
\begin{figure*}[th]
\includegraphics[width=8.5cm, height=5.5cm]{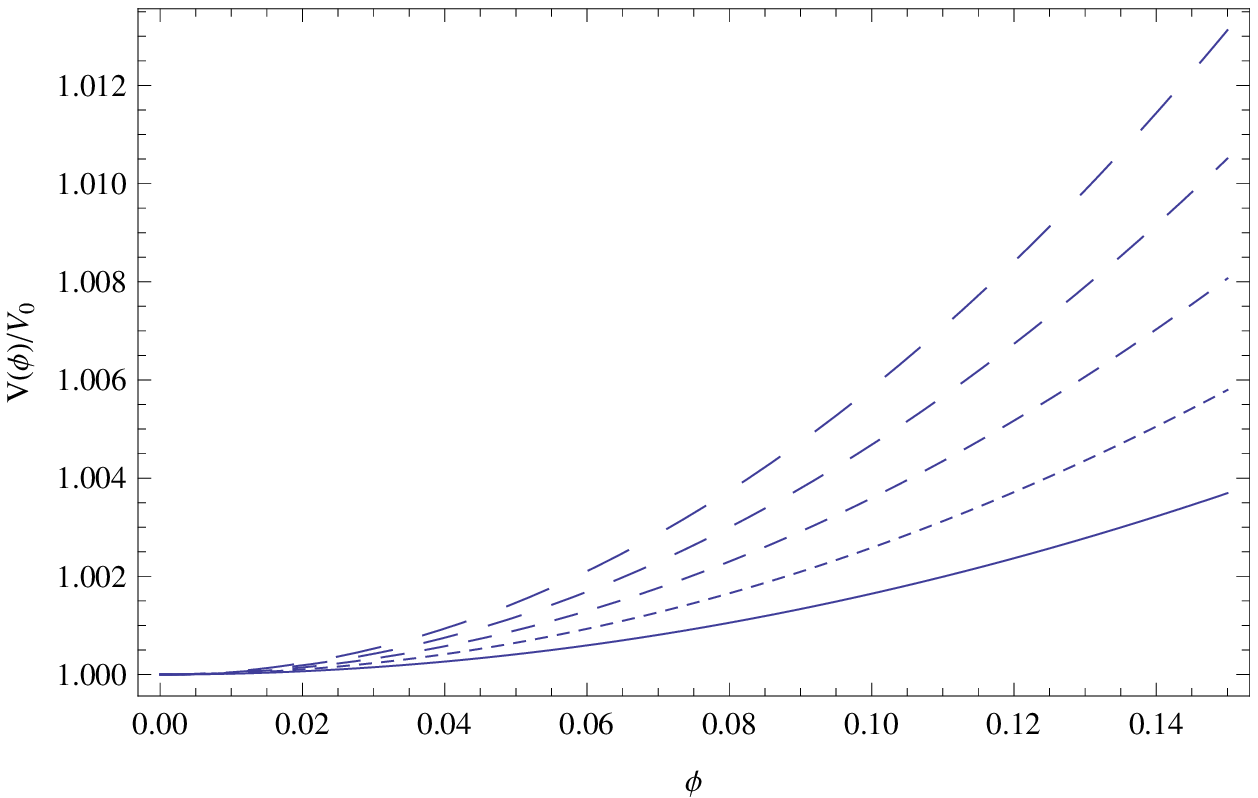}
\includegraphics[width=8.5cm, height=5.5cm]{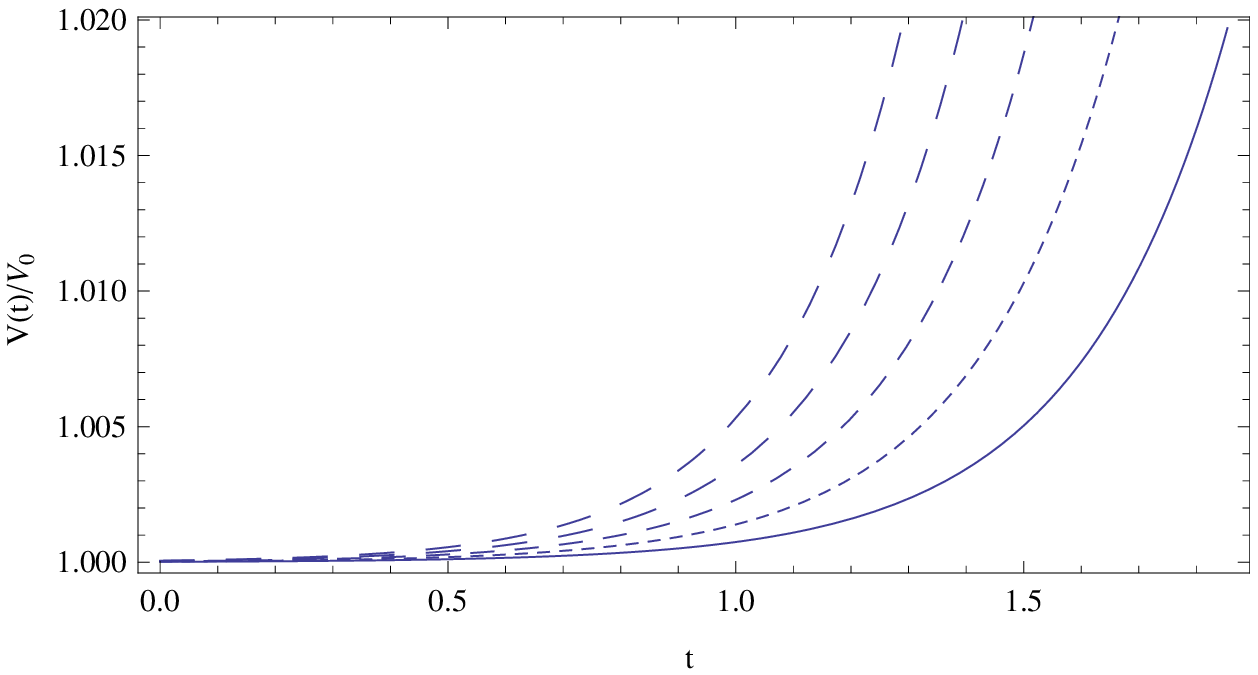}
\caption{Depicted is the variation of the generalized hyperbolic cosine scalar field potential as a function of $\phi $, in the left plot, and as the variation in time, in the right plot, for different values of $\alpha _1$: $\alpha _1=0.1 $ (solid curve), $\alpha _1=0.15$ (dotted curve), $\alpha _1=0.20$ (short dashed curve), $\alpha _1=0.25$ (dashed curve), and $\alpha _1=0.30$ (long dashed curve), respectively.  }\label{hyp1}
\end{figure*}
\end{centering}

\begin{centering}
\begin{figure*}[th]
\includegraphics[width=8.5cm, height=5.5cm]{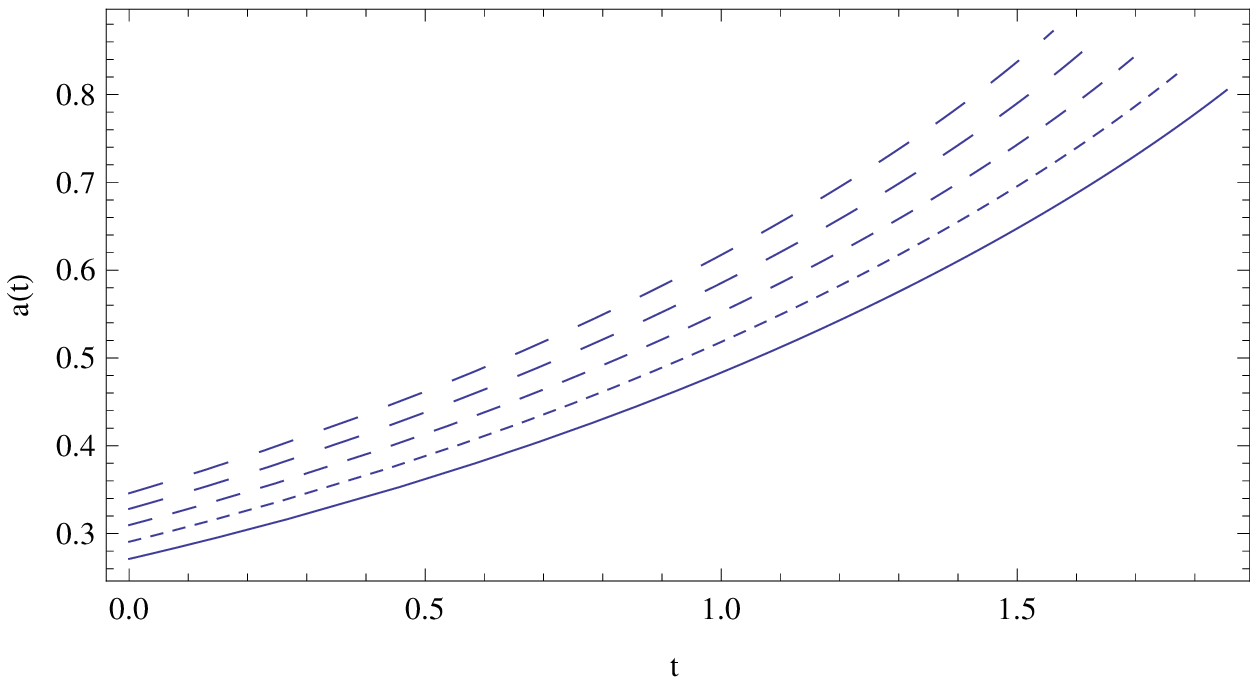}
\includegraphics[width=8.5cm, height=5.5cm]{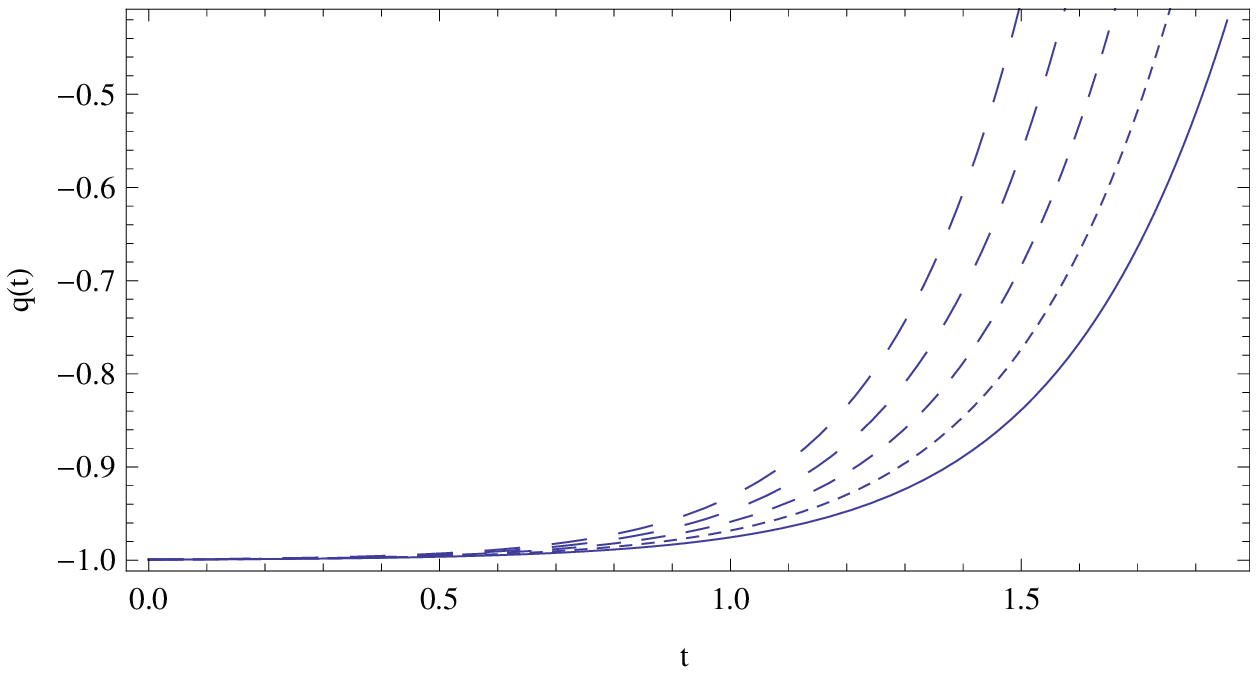}
\caption{Depicted is the time variation of the scale factor, in the left plot, and of the deceleration parameter, in the right plot, of the Universe filled with a scalar field with a generalized hyperbolic cosine self-interaction potential for different values of $\alpha _1$: $\alpha _1=0.1 $ (solid curve), $\alpha _1=0.15$ (dotted curve), $\alpha _1=0.20$ (short dashed curve), $\alpha _1=0.25$ (dashed curve), and $\alpha _1=0.30$ (long dashed curve), respectively.  }
\label{hyp4}
\end{figure*}
\end{centering}

Hence, the exact solution of the field equations for a hyperbolic cosine
type scalar field potential can be obtained in an exact parametric form. The
variations of the scalar field potential as a function of $\phi $ and $t$,
respectively, of the scale factor of the Universe, and of the deceleration
parameter, are represented, for different values of $\alpha _1$, in Figs.~%
\ref{hyp1}-\ref{hyp4}, respectively.

In all of the considered models the Universe shows an expansionary,
accelerated behavior, starting with an initial value $q=-1$ of the
deceleration parameter. The scalar field potential is increasing in time,
leading, in the long time limit, to accelerated expansions, with $q > -1$.

\subsection{Power-law type scalar field potential}

A simple solution of the gravitational field equations for a power-law type
scalar field potential can be obtained by assuming for the function $G$ the
following form
\begin{equation}
G=\mathrm{arccoth}\left( \sqrt{\frac{3}{2}}\frac{\phi }{\alpha _{2}}\right)
,\qquad \alpha _{2}=\mathrm{constant}.
\end{equation}

With this choice of $G$, then Eq.~(\ref{fin}) immediately provides the
scalar field potential given by
\begin{equation}
V\left( \phi \right) =V_{0}\left( \frac{\phi }{\alpha _{2}}\right)
^{-2\left( \alpha _{2}+1\right) }\left[ \frac{3}{2}\left( \frac{\phi }{%
\alpha _{2}}\right) ^{2}-1\right] ,  \label{mmm}
\end{equation}%
where $V_{0}$ is an arbitrary constant of integration. The time dependence
of the scalar field is given by a simple power law,
\begin{equation}
\frac{\phi (t)}{\alpha _{2}}=\left[ \frac{\sqrt{2V_{0}}\left( \alpha
_{2}+2\right) }{\alpha _{2}}\right] ^{\frac{1}{\alpha _{2}+2}}\left(
t-t_{0}\right) ^{\frac{1}{\alpha _{2}+2}}.
\end{equation}%
The scale factor can be obtained from Eq.~(\ref{a}), $da/d\phi=\left[(1/\sqrt{%
6})\coth G\right] a=\left(\phi /2\alpha _{2}\right)a$, and has an explicit exponential dependence
on the scalar field and the time, namely,
\bea
a&=&a_{0}\exp \left( \frac{\phi ^{2}}{4\alpha _{2}}\right) = \nonumber\\
&&a_0\exp \left\{
\frac{1}{4\alpha _{2}}\left[ \frac{\left( \alpha _{2}+2\right) \sqrt{2V_{0}}%
}{\alpha _{2}}\right] ^{\frac{2}{\alpha _{2}+2}}\left( t-t_{0}\right) ^{%
\frac{2}{\alpha _{2}+2}}\right\} ,\nonumber\\
\eea
with $a_{0}$ an arbitrary constant of integration. The deceleration
parameter is given by
\bea
q&=&2\left( \frac{\phi }{\alpha _{2}}\right) ^{-2}-1=
   \nonumber\\
&&2\left[ \frac{\sqrt{2V_{0}%
}\left( \alpha _{2}+2\right) }{\alpha _{2}}\right] ^{-\frac{2}{\alpha _{2}+2}%
}\left( t-t_{0}\right) ^{-\frac{2}{\alpha _{2}+2}}-1.
\eea

\section{Further integrability cases for scalar field cosmologies}
\label{secn}

In the present Section, we will consider several general integrability cases
of Eqs.~(\ref{H}) and (\ref{phi}), describing the dynamics of a scalar field
filled homogeneous and isotropic space-time. By introducing a new set of
variables, the basic equation (\ref{fin}) can be separated into two ordinary
first order differential equations. The resulting compatibility condition
can be integrated exactly for two different forms of the scalar field
potential, thus leading to some exactly integrable classes of the field
equations.

\subsection{The general integrability condition for the field equations}

We rewrite the hyperbolic function $\coth G=\left( 1+e^{-2G}\right) /\left(
1-e^{-2G}\right) $ in the form $\coth G=\left( 1+w^{2}\right) /\left(
1-w^{2}\right) $, where we have introduced a new function $w$ defined as $%
w=e^{-G}$. By substituting $\coth G=\left( 1+w^{2}\right) /\left(
1-w^{2}\right) $, $-(1/w)(dw/d\phi )=dG/d\phi $ into Eq.~(\ref{fin}), and by
denoting $S\left( \phi \right) =-d\ln \left| \sqrt{V}\right| /d\phi $ and $%
\alpha _{3}=-\sqrt{3/2}$, then Eq.~(\ref{fin}) takes the form%
\begin{equation}
\frac{dw}{d\phi }+\left[ \alpha _{3}+S\left( \phi \right) \right] w=\frac{1}{%
3}\frac{dw^{3}}{d\phi }+\left[ \alpha _{3}-S\left( \phi \right) \right]
w^{3}=M\left( \phi \right) ,  \label{n1}
\end{equation}%
where we have introduced a new separation function $M\left( \phi \right) $.
Therefore, we have obtained two linear differential equations for $w$ and $%
w^{3}$, given by
\begin{equation}
\frac{dw}{d\phi }+\left[ \alpha _{3}+S\left( \phi \right) \right] w=M\left(
\phi \right) ,  \label{n2}
\end{equation}%
and
\begin{equation}
\frac{dw^{3}}{d\phi }+3\left[ \alpha _{3}-S\left( \phi \right) \right]
w^{3}=3M\left( \phi \right) ,  \label{n3}
\end{equation}%
respectively. Equation~(\ref{n2}) can be integrated to provide
\begin{equation}
w=\sqrt{V}e^{-\alpha _{3}\phi }\left[ C_{0}+\int \frac{M\left( \phi \right)
e^{\alpha _{3}\phi }}{\sqrt{V}}d\phi \right] ,  \label{n4}
\end{equation}%
where $C_{0}$ is an arbitrary constant of integration. Equation~(\ref{n3})
can be integrated to give
\begin{equation}
w=\frac{e^{-\alpha _{3}\phi }}{\sqrt{V}}\left[ C_{1}+3\int M\left( \phi
\right) V^{3/2}e^{3\alpha _{3}\phi }d\phi \right] ^{1/3},  \label{n5}
\end{equation}%
where $C_{1}$ is an arbitrary constant of integration. Using Eqs.~(\ref{n4})
and (\ref{n5}), we obtain a consistency integral relation between the
separation function $M(\phi )$ and the self-interaction potential $V(\phi )$%
, given by
\begin{equation}
C_{1}+3\int M\left( \phi \right) V^{3/2}e^{3\alpha _{3}\phi }d\phi =V^{3}
\left[ C_{0}+\int \frac{M\left( \phi \right) e^{\alpha _{3}\phi }}{\sqrt{V}}%
d\phi \right] ^{3}.  \label{n7}
\end{equation}

In the following, we shall solve the integral equation (\ref{n7}) for two
particular cases.

\subsubsection{Specific case I: $M\left( \protect\phi \right) =\protect\sqrt{%
V}$}

First we assume that the separation function $M(\phi )$ takes the form $%
M(\phi )=\sqrt{V}$. By substituting $M(\phi )=\sqrt{V}$ into Eq.~(\ref{n7}),
the latter gives an integral equation for the potential $V(\phi )$
\begin{equation}
C_{1}+3\int V^{2}e^{3\alpha _{3}\phi }d\phi =V^{3}A\left( \phi \right) ,
\label{n8}
\end{equation}%
where we have denoted $A(\phi )=\left( C_{0}+e^{\alpha _{3}\phi }/\alpha
_{3}\right) ^{3}$. In order to solve the integral Eq.~(\ref{n8}), we rewrite
it as a linear first order differential equation for $V\left( \phi \right) $
\begin{equation}
\frac{dV}{d\phi }+\left[ \frac{d}{d\phi }\left( \ln A^{1/3}\right) \right] V=%
\frac{e^{3\alpha _{3}\phi }}{A\left( \phi \right) },  \label{n9}
\end{equation}%
with the general solution given by
\begin{equation}
V(\phi )=A^{-1/3}\left( \phi \right) \left[ C_{2}+\int e^{3\alpha _{3}\phi
}A^{-2/3}\left( \phi \right) d\phi \right] ,  \label{nn}
\end{equation}%
where $C_{2}$ is an arbitrary constant of integration. Now by inserting $%
A\left( \phi \right) $ into Eq.~(\ref{nn}) yields the expression of the
scalar field potential as
\begin{widetext}
\begin{equation}
V(\phi )=\frac{\alpha _{3}^{2}\left\{ e^{3\alpha _{3}\phi }+2C_{0}\alpha
_{3}e^{2\alpha _{3}\phi }-C_{0}^{3}\alpha _{3}^{3}+\left( C_{0}\alpha
+e^{\alpha _{3}\phi }\right) ^{2}\left[ \frac{C_{2}}{\alpha _{3}}%
-2C_{0}\alpha _{3}\ln \left| C_{0}\alpha _{3}+e^{\alpha _{3}\phi }\right| %
\right] \right\} }{\left( C_{0}\alpha _{3}+e^{\alpha _{3}\phi }\right) ^{3}}.
\label{pot1}
\end{equation}
\end{widetext}

Therefore we have obtained the following:

\textbf{Theorem 1. } If the scalar field self-interaction potential is given
by Eq.~(\ref{pot1}), then the general solution of Eq.~(\ref{fin}) is given
by
\be
G=\mathrm{arccoth}\left( \frac{ 1+w^{2}} { 1-w^{2}} %
\right) =\ln \left| \frac{1}{w}\right| ,
\ee
where
\be
w(\phi )=\sqrt{V(\phi )}%
e^{-\alpha _{3}\phi }\left( C_{0}+\frac{e^{\alpha _{3}\phi }}{\alpha _{3}}\right) .
\ee

\subsubsection{Specific case II: $M\left( \protect\phi \right) =V^{-3/2}$}

Now we assume that the separation function $M\left( \phi \right) $ takes the
form $M\left( \phi \right) =V^{-3/2}$. By substituting $M\left( \phi \right)
=V^{-3/2}$ into Eq.~(\ref{n7}), the latter gives an integral equation for
the potential $V\left( \phi \right) $,
\begin{equation}
C_{0}+\int \frac{e^{\alpha _{3}\phi }}{V^{2}}d\phi =\frac{\left( C_{1}+\frac{%
1}{\alpha _{3}}e^{3\alpha _{3}\phi }\right) ^{1/3}}{V}.  \label{b1}
\end{equation}%
In order to solve Eq.~(\ref{b1}), we rewrite it as a linear first order
differential equation for $V\left( \phi \right) $
\bea
\frac{dV}{d\phi }+\left[ \frac{d}{d\phi }\ln \left| \frac{1}{\left( C_{1}+%
\frac{1}{\alpha _{3}}e^{3\alpha _{3}\phi }\right) ^{1/3}}\right| \right] V
   \nonumber\\
=-\frac{e^{\alpha _{3}\phi }}{\left( C_{1}+\frac{1}{\alpha _{3}}e^{3\alpha
_{3}\phi }\right) ^{1/3}}.  \label{b2}
\eea

Equation~(\ref{b2}) can be easily integrated, and yields the following
solution
\bea
V\left( \phi \right) &=&\left( C_{1}+\frac{1}{\alpha _{3}}e^{3\alpha _{3}\phi
}\right) ^{1/3}\times\nonumber\\
&&\times \left[ C_{3}-\int \frac{e^{\alpha _{3}\phi }}{\left( C_{1}+%
\frac{1}{\alpha _{3}}e^{3\alpha _{3}\phi }\right) ^{2/3}}d\phi \right] .
\label{pot2}
\eea%
where $C_{3}$ is an arbitrary constant of integration.

Therefore we have obtained the following:

\textbf{Theorem 2. } If the scalar field self-interaction potential is given
by Eq.~(\ref{pot2}), then the general solution of Eq.~(\ref{fin}) is given
by
\be
G=\mathrm{arccoth}\left( \frac{ 1+w^{2}}{1-w^{2}} %
\right) =\ln \left| \frac{1}{w}\right| ,
\ee
 where
 \be
 w(\phi )=\left[ \frac{e^{-\alpha
_{3}\phi }}{\sqrt{V(\phi )}}\right] \left( C_{1}+\frac{e^{3\alpha _{3}\phi }}{\alpha
_{3}}\right) ^{1/3}.
\ee

\section{Approximate solutions for scalar fields with arbitrary
self-interaction potentials}
\label{sect5}

Generally, for a given scalar field potential $V=V(\phi )$, Eq. (\ref{fin})
cannot be solved exactly. However, there are several limiting cases in which
approximate general solutions can be obtained for arbitrary potentials.
These cases correspond to the approximation of the function
\begin{equation}
\coth G(\phi )=\sqrt{1+\frac{2V}{\dot{\phi}^{2}}},  \label{14}
\end{equation}%
in the asymptotic limits of small and large $G$ by some simple analytical
expressions.

\subsection{The limit of large $G$}

One such important case is the limit of large $G$, $G\rightarrow \infty $,
when $\coth G=1$. From Eq.~(\ref{14}) it follows that this case corresponds
to scalar fields satisfying the condition $\dot{\phi}^{2}/2\gg V$.
Therefore, Eq. (\ref{fin}) takes the form
\begin{equation}
\frac{dG}{d\phi }+\frac{1}{2V}\frac{dV}{d\phi }+\sqrt{\frac{3}{2}}=0,
\end{equation}%
with the general solution given by
\begin{equation}
G\left( \phi \right) =C_{4}-\sqrt{\frac{3}{2}}\phi -\frac{1}{2}\ln \left|
V(\phi )\right| ,
\end{equation}%
where $C_{4}$ is an arbitrary constant of integration. Then we obtain the
following solutions
\begin{eqnarray}
u(\phi ) &=&\cosh \left[ C_{4}-\sqrt{\frac{3}{2}}\phi -\frac{1}{2}\ln \left|
V(\phi )\right| \right] , \\
F(\phi ) &=&\sqrt{V(\phi )}\cosh \left[ C_{4}-\sqrt{\frac{3}{2}}\phi -\frac{1%
}{2}\ln \left| V(\phi )\right| \right],
\end{eqnarray}%
respectively.

By using the relation between $\dot{\phi}$ and $F$, $\dot{\phi}=\sqrt{2\left[
F^{2}(\phi )-V(\phi )\right] }=\sqrt{2V(\phi )}\sinh G(\phi )$, we obtain
\begin{equation}
t-t_{0}=\frac{1}{\sqrt{2}}\int \frac{d\phi }{\sqrt{V\left( \phi \right) }%
\sinh \left[ C_{4}-\sqrt{\frac{3}{2}}\phi -\frac{1}{2}\ln \left| V(\phi
)\right| \right] }.  \label{case1t}
\end{equation}%
The scale factor $a$ can be obtained as
\begin{equation}
\ln \left| \frac{a}{a_{0}}\right| =\frac{1}{\sqrt{6}}\int \coth \left[ C_{4}-%
\sqrt{\frac{3}{2}}\phi -\frac{1}{2}\ln \left| V(\phi )\right| \right] d\phi ,
\label{case1a}
\end{equation}%
where $a_{0}$ is an arbitrary constant of integration. As one can see from
Eq.~(\ref{case1a}), in the limit of large $G$, $\coth \left[ C_{4}-\sqrt{%
\frac{3}{2}}\phi -\frac{1}{2}\ln \left| V(\phi )\right| \right] \rightarrow
1 $, the scale factor can be represented as an exponential function of the
scalar field,
\begin{equation}
a=a_{0}\exp \left( \frac{\phi }{\sqrt{6}}\right) .
\end{equation}%
Equations~(\ref{case1t}) and (\ref{case1a}) give a parametric representation
of the time variation of the scale factor, with the scalar field $\phi $
taken as parameter. The deceleration parameter $q$ is given in this limit by
\begin{equation}
q=3\tanh ^{2}\left[ C_{4}-\sqrt{\frac{3}{2}}\phi -\frac{1}{2}\ln \left|
V(\phi )\right| \right] -1.
\end{equation}%
In the limit of large $G$, corresponding to the limit of $\tanh ^{2}\left[
C_{4}-\sqrt{\frac{3}{2}}\phi -\frac{1}{2}\ln \left| V(\phi )\right| \right]
\rightarrow 1$, we have $q\approx 2$.

\subsection{The limit of small G}

A second case in which an approximate general solution of Eq. (\ref{fin})
can be found for arbitrary potentials corresponds to the limit of small $G$,
$G\rightarrow 0$, when $\coth G\rightarrow \infty $. This condition is
satisfied for potential dominated scalar fields, with $\dot{\phi}^{2}/2\ll V$%
. In this case, one can neglect the small term of the order of unity $\sqrt{%
3/2}$ in the equation (\ref{fin}), thus obtaining
\begin{equation}
\frac{dG}{d\phi }+\frac{1}{2V}\frac{dV}{d\phi }\coth G=0.  \label{appr2}
\end{equation}%
The general solution of Eq. (\ref{appr2}) is
\begin{equation}
G=\mathrm{arccosh}\left( \frac{C_{5}}{\sqrt{V}}\right) ,
\end{equation}%
where $C_{5}>0$ is an arbitrary constant of integration.

As in the previous case we obtain
\begin{equation}
u=\frac{C_{5}}{\sqrt{V}},\qquad F=C_{5}.
\end{equation}%
This gives immediately
\begin{equation}
t-t_{0}=\frac{1}{\sqrt{2}}\int \frac{d\phi }{\sqrt{C_{5}^{2}-V}},
\end{equation}%
and
\begin{equation}
a=a_{0}\exp \left[ \frac{C_{5}}{\sqrt{3}}\left( t-t_{0}\right) \right] .
\end{equation}

The deceleration parameter is obtained as
\begin{equation}
q=2-\frac{3V}{C_{5}^{2}}.
\end{equation}

\subsection{Power series solution of the field equations}

The hyperbolic function $\coth G$ is given by
\bea
\coth G&=&\frac{1}{G}+\frac{G}{3%
}-\frac{G^{3}}{45}+\frac{2G^{5}}{945}...
   \nonumber\\
&=&\frac{1}{G}+\sum_{n=1}^{\infty }%
\frac{2^{2n}B_{2n}G^{2n-1}}{\left( 2n\right) !}, \quad 0<\left| G\right| <\pi ,
\eea
where we have introduced the $n^{th}$ Bernoulli number $B_{2n}$, defined in
terms of the Riemann zeta function, and  given by  \cite{PoZa}
\be
B_{2n}=\left( -1\right) ^{n+1}%
\frac{2\left( 2n\right) !}{\left( 2\pi \right) ^{2n}}\left( 1+\frac{1}{2^{2n}%
}+\frac{1}{3^{2n}}+\frac{1}{4^{2n}}+...\right).
\ee

Now, we consider the limit of small $G$ in this series expansion, and
restrict the series expansion to the first two terms, so that $\coth
G\approx 1/G+G/3$. Then Eq.~(\ref{fin}) takes the form
\begin{equation}
G\frac{dG}{d\phi }+\frac{1}{6V}\frac{dV}{d\phi }G^{2}+\sqrt{\frac{3}{2}}G+%
\frac{1}{2V}\frac{dV}{d\phi }=0.  \label{kk}
\end{equation}%
By means of the transformation $h=1/G$, Eq.~(\ref{kk}) becomes a first order
Abel equation given by
\begin{equation}
\frac{dh}{d\phi }-\frac{1}{2V}\frac{dV}{d\phi }h^{3}-\sqrt{\frac{3}{2}}h^{2}-%
\frac{1}{6V}\frac{dV}{d\phi }h=0.  \label{abel1}
\end{equation}
Then, by introducing a new variable $\eta $ by means of the transformation $%
h=V^{1/6}(\phi )\eta (\phi) $, we obtain the equation
\begin{equation}
\frac{d\eta }{d\phi }-\frac{1}{2V^{2/3}}\frac{dV}{d\phi }\eta ^{3}-\sqrt{%
\frac{3}{2}}V^{1/6}\eta ^{2}=0.  \label{abel2}
\end{equation}
A change of the independent variable $\phi $ to $\xi =\sqrt{3/2}\int
V^{1/6}d\phi $ leads to
\begin{equation}
\frac{d\eta }{d\xi }-\frac{1}{2V^{2/3}}\frac{dV}{d\xi }\eta ^{3}-\eta ^{2}=0.
\label{eta}
\end{equation}

By taking $\eta =-(1/\psi )(d\psi /d\xi )$, with the use of the mathematical
identity $d^{2}\xi /d\psi ^{2}=-\left( d^{2}\psi /d\xi ^{2}\right) /\left(
d\psi /d\xi \right) ^{3}$, we obtain the following second order equation for
$\psi $,
\begin{equation}
\psi ^{2}\frac{d^{2}\xi }{d\psi ^{2}}+\frac{1}{2V^{2/3}}\frac{dV}{d\xi }=0.
\label{abelfin}
\end{equation}

In the following we denote $2^{-1}V^{-2/3}dV/d\xi =(3/2)d\left(
V^{1/3}\right) /d\xi =\chi \left( \xi \right) $. By introducing the
transformations $\xi =\psi \sigma $ and $\tau =1/\psi $, we obtain for Eq.~(%
\ref{abelfin}) the form
\begin{equation}
\tau \frac{d^{2}\sigma }{d\tau ^{2}}+\chi \left( \frac{\sigma }{\tau }%
\right) =0.  \label{abelfin1}
\end{equation}

\subsection{Exact integrable scalar field potentials}

Equation (\ref{abelfin}) can be integrated exactly in several cases. In the
first case
\begin{equation}
\frac{1}{2V^{2/3}}\frac{dV}{d\xi }=\frac{3}{2}m_{0}  \label{mm}
\end{equation}
where $m_{0}$ is an arbitrary constant, Eq.~(\ref{mm}) can be easily
integrated to give
\begin{equation}
V^{1/3}=m_{0}\xi .
\end{equation}%
Taking into account the definition of $\xi $, we obtain immediately for $V$
the functional form $V(\phi )\sim \phi ^{6}$. Another case of integrability
of Eq.~(\ref{abelfin}) corresponds to
\begin{equation}
\frac{1}{2V^{2/3}}\frac{dV}{d\xi }=3m_{0}\xi ,
\end{equation}%
giving the case of the exponential potential, $V(\phi )\sim \exp (\sqrt{%
3m_{0}/2}\phi )$.

As a last case of exactly integrable scalar field models, we consider a
scalar field potential of the form
\begin{equation}
V(\phi) =\left(\sqrt{\frac{2}{3}}\right)^6V_0^6\cosh ^6\left(\phi-\phi
_0\right),
\end{equation}
where $V_0$ and $\phi _0$ are constants. Then we obtain first
\begin{equation}
\xi =V_0\sinh \left(\phi -\phi _0\right),
\end{equation}
and
\begin{equation}
V(\xi)=\left(\sqrt{\frac{2}{3}}\right)^6 V_0^6\left(1+\frac{\xi ^2}{V_0^2}%
\right)^3.
\end{equation}

Equation~(\ref{abelfin}) becomes
\begin{equation}
\psi ^{2}\frac{d^{2}\xi }{d\psi ^{2}}+2\xi =0,
\end{equation}
with the general solution given by
\begin{equation}
\xi (\psi )=\sqrt{\psi }\left[\xi_1 \sin \left(\frac{\sqrt{7}}{2} \ln |\psi|
\right)+\xi_2 \cos \left(\frac{\sqrt{7}}{2} \ln |\psi |\right)\right],
\end{equation}
where $\xi _1$ and $\xi _2$ are arbitrary constants of integration. For $%
\eta $ we obtain
\begin{widetext}
\begin{equation}
\eta =-\frac{1}{\psi \left(d\xi /d\psi \right)}=-\frac{2}{\sqrt{\psi } \left[%
\left(\xi_1-\sqrt{7} \xi _2\right) \sin \left(\frac{\sqrt{7}}{2} \log |\psi
|\right)+\left(\sqrt{7} \xi _1+\xi_2\right) \cos \left(\frac{\sqrt{7}}{2}
\log |\psi |\right)\right]},
\end{equation}
giving
\begin{equation}
G=\frac{1}{h}=\frac{1}{V^{1/6}(\psi )\eta (\psi)}=-\frac{\sqrt{\frac{3}{2}}
\sqrt{\psi } \left[\left(\xi _1-\sqrt{7} \xi _2\right) \sin \left(\frac{%
\sqrt{7}}{2} \ln |\psi |\right)+\left(\sqrt{7} \xi _1+\xi _2\right) \cos
\left(\frac{\sqrt{7}}{2} \ln |\psi |\right)\right]}{2 \sqrt{\psi \left[\xi
_1 \sin \left(\frac{\sqrt{7}}{2} \ln |\psi | \right)+\xi _2 \cos \left(\frac{%
\sqrt{7}}{2} \ln |\psi | \right)\right]^2+V_0^2}}.
\end{equation}
\end{widetext}

With this expression of $G$, the solution of the field equations for the $%
\cosh ^{6}\left( \phi -\phi _{0}\right) $ scalar field potential in the
intermediate regime can be obtained in an exact parametric form, with $\psi $
taken as parameter.

\section{The simple power law scalar field potential}
\label{sect6}

A class of scalar field potentials which have been extensively considered in
the physical literature as a possible potential for the inflaton field is
the simple power law potential \cite{pow},
\begin{equation}
V=V_{0}\phi ^{\sqrt{6}\lambda },
\end{equation}
with $V_{0}$ and $\lambda $ constants. This class of potentials includes the
simplest chaotic models, in which inflation starts from large values for the
inflaton, with inflation ending by violating the slow-roll regime \cite{P1}.
The model with a quadratic potential, $\lambda = 2/\sqrt{6}$, is considered
the simplest example of inflation. For this class of potentials, Eq.~(\ref%
{fin}) takes the form
\begin{equation}
\frac{dG}{d\phi }+\sqrt{\frac{3}{2}}\frac{\lambda }{\phi }\coth G+\sqrt{%
\frac{3}{2}}=0.  \label{po1}
\end{equation}

We consider below several cases of interest.

\subsection{The solution of the field equations in the large and small limit
of $G$}

The solution of Eq.~(\ref{po1}) can be immediately obtained in the
asymptotic limit of large and small $G$, respectively. For $G\rightarrow
\infty $, $\coth G\rightarrow 1$, we obtain
\begin{equation}
G\left( \phi \right) =\sqrt{\frac{3}{2}}\left( C_{6}-\phi -\lambda \ln
\left| \phi \right| \right) ,
\end{equation}%
where $C_{6}$ is an arbitrary constant of integration. Thus, in this limit
the general solutions of the field equations are the following
\begin{equation}
t-t_{0}=\frac{1}{\sqrt{2V_{0}}}\int \frac{\phi ^{-\sqrt{\frac{3}{2}}\lambda
}d\phi }{\sinh \left[ \sqrt{\frac{3}{2}}\left( C_{6}-\phi -\lambda \ln
\left| \phi \right| \right) \right] },
\end{equation}%
\begin{equation}
\ln \left| \frac{a}{a_{0}}\right| =\frac{1}{\sqrt{6}}\int \coth \left[ \sqrt{%
\frac{3}{2}}\left( C_{6}-\phi -\lambda \ln \left| \phi \right| \right) %
\right] d\phi ,
\end{equation}%
and
\begin{equation}
q=3\tanh ^{2}\left[ \sqrt{\frac{3}{2}}\left( C_{6}-\phi -\lambda \ln \left|
\phi \right| \right) \right] -1,
\end{equation}%
respectively.

In the opposite limit $G\rightarrow 0$, we obtain the equation
\begin{equation}
\frac{dG}{d\phi }+\sqrt{\frac{3}{2}}\frac{\lambda }{\phi }\coth G=0,
\end{equation}%
which provides
\begin{equation}
\cosh G=C_{6}\phi ^{-\sqrt{\frac{3}{2}}\lambda }.
\end{equation}

Therefore the general solutions of the field equations are given by
\begin{eqnarray}
t-t_{0} &=&\frac{1}{\sqrt{2V_{0}}}\int \frac{d\phi }{\sqrt{C_{6}^{2}-\phi ^{2%
\sqrt{\frac{3}{2}}\lambda }}}, \\
\ln \left| \frac{a}{a_{0}}\right| &=&\frac{C_{6}}{\sqrt{6}}\int \frac{d\phi
}{\sqrt{C_{6}^{2}-\phi ^{2\sqrt{\frac{3}{2}}\lambda }}},
\end{eqnarray}%
and
\begin{equation}
q=2-\frac{3}{C_{6}^{2}}\phi ^{2\sqrt{\frac{3}{2}}\lambda },
\end{equation}%
respectively.

\subsection{The intermediate regime}

In the intermediate regime the dynamics of the power law potential scalar
field filled Universe is described by Eq.~(\ref{abelfin1}). In the following
it is more convenient to re-scale the potential so that $\sqrt{6}\lambda
\rightarrow \lambda $. Hence the potential can be written as $V=V_{0}\phi
^{\lambda }$.

To obtain the actual form of Eq.~(\ref{abelfin1}) we find %
%
\begin{equation}
\xi = \sqrt{\frac{3}{2}}\int V^{1/6}d\phi =\sqrt{\frac{3}{2}}\;V_{0}^{1/6}
\frac{\phi ^{\left( \lambda +6\right) /6}}{\left( \lambda /6+1\right)} \,,
\end{equation}
giving
\begin{equation}
\phi =\left[ \sqrt{\frac{2}{3}}\left( \frac{\lambda}{6}+1\right) \right]
^{6/(\lambda +6)}V_{0}^{-1/(\lambda +6)}\;\xi ^{6/\left( \lambda +6\right)
}\,,
\end{equation}
and
\begin{equation}
V\left( \xi \right) =V_{0\xi }\xi ^{6\lambda /\left( \lambda +6\right) } \,,
\end{equation}
where
\begin{equation}
V_{0\xi }=\left[ \sqrt{\frac{2}{3}}\left( \frac{\lambda}{6}+1\right) \right]
^{6\lambda /(\lambda +6)}V_{0}^{6/(\lambda +6)} \,.
\end{equation}

For the function $\chi \left( \xi \right) $ we find
\begin{equation}
\chi \left( \xi \right)
=\left( 3V_{0\xi }^{1/3}/2\right) d\xi ^{2\lambda /\left( \lambda +6\right)
}/d\xi =\chi _{0}\xi ^{\left( \lambda -6\right) /\left( \lambda +6\right) }\,,
\end{equation}
with $\chi _{0}=\left[ 3\lambda /\left( \lambda +6\right) \right] V_{0\xi
}^{1/3}$. Therefore for the simple power-law potential, Eq. (\ref{abelfin1})
takes the form
\begin{equation}
\frac{d^{2}\sigma }{d\tau ^{2}}+\chi _{0}\tau ^{-\frac{2\lambda }{\lambda +6}%
}\sigma ^{\frac{\lambda -6}{\lambda +6}}=0.  \label{emden}
\end{equation}

Equation~(\ref{emden}) can be immediately integrated for the case $\lambda
=6 $, that is, for a scalar field self interaction potential of the form $%
V=V_{0}\phi ^{6}$. In this case Eq.~(\ref{emden}) becomes
\begin{equation}
\frac{d^{2}\sigma }{d\tau ^{2}}+\frac{\chi _{0}}{\tau }=0,
\end{equation}%
with the general solution given by
\begin{equation}
\sigma \left( \tau \right) =-\chi _{0}\tau \ln \left| \tau \right|
+C_{7}\tau +C_{8},
\end{equation}%
where $C_{7}$ and $C_{8}$ are arbitrary constants of integration.

The successive transformations
\begin{eqnarray}
\sigma \left( \psi \right) &=&\chi _{0}\ln
\left| \psi \right| /\psi +C_{7}/\psi +C_{8} ,\\
 \xi &=&\psi \sigma =\chi
_{0}\ln \left| \psi \right| +C_{8}\psi +C_{7},\\
 \eta &=&\left( 1/\psi \right)
\left( 1/d\xi /d\psi \right) =1/\left( C_{8}\psi +\chi _{0}\right),
\end{eqnarray}
and $G=\psi \left( d\xi /d\psi \right) V^{-1/6}$, yield
\begin{equation}
G=V_{0\xi }^{-1/6}\frac{C_{8}\psi +\chi _{0}}{\left( \chi _{0}\ln \left|
\psi \right| +C_{8}\psi +C_{7}\right) ^{1/2}},
\end{equation}%
where for $\lambda =6$, $V_{0\xi }=\left( 2\sqrt{2/3}\right) ^{3}V_{0}^{1/2}$.

The knowledge of $G$ allows us to obtain immediately the general solution of
the field equations in the intermediary regime in the following parametric
form, with $\psi $ taken as parameter:
\begin{widetext}
\begin{equation}
t-t_{0}=\frac{\sqrt{3}}{8}V_{0}^{-1/3}\int \frac{1}{\sinh \left[ V_{0\xi
}^{-1/6}\frac{C_{8}\psi +\chi _{0}}{\left( \chi _{0}\ln \left| \psi \right|
+C_{8}\psi +C_{7}\right) ^{1/2}}\right] }\frac{\left( \frac{\chi _{0}}{\psi }%
+C_{8}\right) }{\left( \chi _{0}\ln \left| \psi \right| +C_{8}\psi
+C_{7}\right) ^{2}}d\psi ,
\end{equation}
\begin{equation}
\phi =\frac{2}{6^{1/4}}\frac{1}{V_{0}^{1/12}}\left( \chi _{0}\ln \left| \psi
\right| +C_{8}\psi +C_{7}\right) ^{1/2},  \label{AAA}
\end{equation}%
\begin{equation}
\ln \left| \frac{a}{a_{0}}\right| =-\frac{1}{6^{3/4}V_{0}^{1/12}}\int \coth %
\left[ V_{0\xi }^{-1/6}\frac{C_{8}\psi +\chi _{0}}{\left( \chi _{0}\ln
\left| \psi \right| +C_{8}\psi +C_{7}\right) ^{1/2}}\right] \frac{\left(
\frac{\chi _{0}}{\psi }+C_{8}\right) }{\left( \chi _{0}\ln \left| \psi
\right| +C_{8}\psi +C_{7}\right) ^{1/2}}d\psi ,
\end{equation}%
\end{widetext}
and
\begin{equation}
q=3\tanh ^{2}\left[ V_{0\xi }^{-1/6}\frac{C_{8}\psi +\chi _{0}}{\left( \chi
_{0}\ln \left| \psi \right| +C_{8}\psi +C_{7}\right) ^{1/2}}\right] -1,
\end{equation}
respectively.

\section{Discussions and final remarks}

\label{sect7}

In the present paper, we have considered a systematic approach for the study
of scalar field cosmological models in a flat, homogeneous and isotropic
space-time. With the help of some simple transformations, and the use of the
gravitational field equations, the Klein-Gordon equation describing the
dynamics of the scalar field can be transformed to a first order non-linear
differential equation for the new unknown function $G$. This equation
immediately leads to the identification of some classes of scalar field
potentials for which the field equations can be solved exactly, and it
allows the formulation of general integrability conditions. In this context,
we have obtained the general solutions of the gravitational field equations
for the cases of the exponential, generalized hyperbolic cosine, and the
generalized power law potentials. Moreover, it can be used to obtain some
simple analytical solutions in the limits of small and large values of the
cosmological parameters, as well as in the intermediate regime.

As a first application of the developed method, we have analyzed the problem
of the exponential potential, which has been previously intensively
investigated in the literature, with several methods being used to obtain
the solution of the cosmological field equations. A Hamilton-Jacobi approach
was proposed in \cite{Sal}, by using a spatial gradient expansion based on
the Arnowitt-Deser-Misner (ADM) formulation of Einstein and scalar-field
equations. By neglecting the second-order spatial gradients, the ADM and
scalar-field equations reduce to the simple collection of background-field
equations
\be
H=H(\phi),
\ee
\be
H^2(\phi)=\frac{m_{P}^2}{12\pi}\left(\frac{\partial
H}{\partial \phi }\right)^2+\frac{8\pi }{3m_P^2}V(\phi),
\ee
\be
\frac{\dot{\phi }}{N}=-\frac{m_P^2}{4\pi}\frac{\partial H}{\partial \phi },
\ee
and $\dot{\alpha }/N=H$, respectively, where $H$ is the Hubble function, $m_P$ is
the Planck mass, $N$ is the lapse function, and $\alpha $ is related to the
scale factor $a$ by $a\left(t,x^j\right)=\exp\left[\alpha \left(t,x^j\right)%
\right]$. For the case of an exponential potential of the form $%
V(\phi)=V_0\exp\left(-\sqrt{16\pi/p}\phi /m_P\right)$, where $p$ is a
constant, by introducing a new dependent variable $f(\phi)$ so that
\be
H(\phi)=%
\sqrt{\frac{8\pi V_0}{3m_P^2}}\exp\left(-\sqrt{\frac{16\pi}{p}}\frac{\phi }{m_P}\right)f(\phi),
\ee
 it
turns out that the function $f$ satisfies the differential equation
\be
\frac{m_P}{12\pi }\frac{df}{d\phi} =\frac{f}{\sqrt{3p}}\pm\left(f^2-1\right),
\ee
which
can be solved by introducing the change of variables $f=\cosh (u)$.
Therefore, the general solution of the field equations for the exponential
scalar field potential is obtained in a parametric form as
\be
H(u)=\sqrt{\frac{8\pi
V_0}{3m_P^2}}\exp\left(-\sqrt{\frac{16\pi}{p}}\frac{\phi}{m_P}\right)\cosh (u),
\ee
 and
 \bea
 \phi(u)&=&\phi _m-\left(\frac{m_P}{\sqrt{12\pi}}\right)\left(1-\frac{1}{3p}\right)^{-1}\Bigg[%
u+(3p)^{-1/2}\times \nonumber\\
&&\ln\left|\cosh (u)-\sqrt{3p}\sinh (u)\right|\Bigg].
\eea
The
parametric representation for $\phi $ obtained in \cite{Sal} is the same as
the one given by Eq.~(\ref{expphi}). However, the time dependence of the
cosmological parameters for the exponential potential case is not discussed
in \cite{Sal}.

Recently, in \cite{K1} and \cite{K2} the cosmological evolution of the scalar field with an exponential potential of
the form $V=V_0\exp(\lambda \phi)$, $V_0,\lambda=\mathrm{constants}$, was
investigated by considering a new time variable. As a starting point two new
variables $u$ and $v$ are introduced via the transformations $a^3=e^{v+u}$
and $\phi =A\left(v-u\right)$, where $A$ is a constant. The Friedmann and
the Klein-Gordon equations take the form
\be
\frac{1}{9}\left(\dot{u}^2+\dot{v}^2+2%
\dot{u}\dot{v}\right)=\frac{A^2}{2}\left(\dot{u}^2+\dot{v}^2-2\dot{u}%
\dot{v}\right)+V_0e^{\lambda A(u-v)},
\ee
and
\be
A\left(\ddot{v} - \ddot{u}%
\right) + A\left(\dot{v}^2 - \dot{u}^ 2\right) + \lambda V_0e^{\lambda
A(v-u)} = 0,
\ee
respectively. By taking $A=\sqrt{2}/3$, the Friedmann equation
becomes
\be
\dot{u}\dot{v}=\frac{9V_0}{4}e^{(\sqrt{2}/3)\lambda (u-v)}.
\ee

In order to further simplify the formalism, a new time variable $\tau $ is
introduced, so that the previous equation becomes
\be
u^{\prime}v^{\prime}\dot{%
\tau }^2=\frac{9V_0}{4}e^{(\sqrt{2}/3)\lambda (u-v)},
\ee
 where prime
denotes the derivative with respect to $\tau $. The new time parameter is
chosen so that
\be
\dot{\tau}=\frac{3}{2}\sqrt{V_0}e^{\lambda \phi /2}=\frac{3}{2}\sqrt{V_0}%
e^{\lambda (u-v)/\sqrt{6}},
\ee
 a choice that simplifies the Friedmann equation
to $u^{\prime}v^{\prime}=1$. By introducing a new variable $x=v^{\prime}$,
the Klein-Gordon equation can be transformed to a Riccati type equation,
\be
x^{\prime}+\left(1+\frac{\sqrt{2}\lambda }{6}\right)x^2+\left(\frac{\sqrt{2}\lambda }
{6}-1\right)=0,
\ee
which by means of the transformation $x=\left[\left(1/1+%
\sqrt{2}\lambda /6\right)\right]y^{\prime}/y$ is transformed to a second
order linear differential equation of the form
\be
y^{\prime\prime}+\left(%
\frac{\lambda ^2}{18}-1\right)y=0.
\ee
Two cases are considered in detail, the
hyperbolic cosine, when the constant $|\lambda |<3\sqrt{2}$, and the
``trigonometric'' case, with $|\lambda |>3\sqrt{2}$, and the corresponding
cosmological dynamics is studied in detail in the new time variable $\tau $.

As compared to the previous studies, the method introduced in the present
paper for the exponential potential scalar field allows the direct study of
the time dependence of the physical parameters of the cosmological models,
without the need of introducing a new time variable. A number of exact
analytical solutions can be obtained in a parametric form from the general
integral representation of the time variable for some specific values of the
coefficient $\alpha _0$. Moreover, by using the exact solutions the limiting
behaviors of the solutions, corresponding to the long time behavior, and
near $t=0$ can be easily obtained. It is also a simpler method, since the
gravitational field equations can be reduced to a first order differential
equation. Once the solution of this basic differential equation is known,
all the physical/cosmological parameters can be obtained in a
straightforward way.

As a second exactly integrable case that can be easily studied with the
present formalism we have considered the generalized cosine hyperbolic
potential given by Eq.~(\ref{kkk}), $V(\phi )=V_{0}\cosh ^{2\alpha
_{1}/\left( 1+\alpha _{1}\right) }\left[ \sqrt{3/2}\left( 1+\alpha
_{1}\right) \left( \phi -\phi _{0}\right) \right] $. Integrable scalar field
models with potential
\be
V(\phi )=C_{1}\left[ \cosh (\gamma \phi )\right]
^{2/\gamma -2}+C_{2}\left[ \sinh (\gamma \phi )\right] ^{2/\gamma -2},
\ee
were
discussed in \cite{Fre1}, where mechanical systems for the $(A,\phi )$
variables, whose equations of motion follow from the class of Lagrangians of
the form
\be
L=e^{A-B}\left[ -\frac{\dot{A}^{2}}{2}+\frac{\dot{\phi}^{2}}{2}-e^{2B}V(\phi )%
\right],
\ee
 were analyzed. By using the formalism introduced in the present
paper the general solution of the gravitational field equations can be
easily obtained.

As a third case of exactly integrable scalar field models we have analyzed
in detail the cosmological dynamics for a Universe filled with a generalized
power law scalar field potential of the form given by Eq. (\ref{mmm}) $%
V\left( \phi \right) =V_{0}\left( \frac{\phi }{\alpha _{2}}\right)
^{-2\left( \alpha _{2}+1\right) }\left[ \frac{3}{2}\left( \frac{\phi }{%
\alpha _{2}}\right) ^{2}-1\right] $, which consists of the sum of two simple
power law potentials, and represents the generalization of the simple power
law potential extensively discussed in \cite{pow}. We have also analyzed in
detail potentials of the form $V=V_{0}\phi ^{\sqrt{6}\lambda }$. In this
case, the general solution of the field equations cannot be obtained in an
exact form, but the limiting small and large time behavior, as well as the
study of the intermediate phases can be performed relatively easily. Two
general integrability conditions for the basic first order differential
equation have also been obtained, corresponding to a given form of the
scalar field potential, given by Eqs.~(\ref{pot1}) and (\ref{pot2}). Such
potentials have not been previously considered in the study of cosmological
scalar field models. However, despite their apparent complexity, the
corresponding gravitational field equations can be solved exactly.

In concluding, we have obtained several exact solutions of the gravitational
field equations whose background cosmological evolutions can reproduce the
results of the standard $\Lambda $CDM cosmological model. In order to obtain
a deeper physical understanding of the solutions a comparison with the
supernovae data is necessary \cite{acc}. In addition to this, in order to
compare the obtained models with the data on the microwave background cosmic
radiation and the large scale structure of the Universe, the study of the
cosmological perturbations of the solutions is necessary in the obtained
theoretical framework. Work under these lines is presently underway, and the
results will be presented in a future publication.

\section*{Acknowledgments}

We would like to thank the anonymous referee for comments and suggestions that helped us to significantly improve our manuscript.  We also thank Dr. Vitaliy
Cherkaskiy, Dr. Yuri Bolotin, Dr. Oleg Lemets and Dr. Danylo Yerokhin for
pointing out an important sign error in the first version of our manuscript.
We also thank Professor Jos\'{e} P. Mimoso for suggestions that helped us to
improve our manuscript. MKM would like to dedicate this paper to his teacher
Professor C. W. Kilmister from King's College London. FSNL is supported by a
Funda\c{c}\~{a}o para a Ci\^{e}ncia e Tecnologia Investigador FCT Research
contract, with reference IF/00859/2012, funded by FCT/MCTES (Portugal). FSNL
also acknowledges financial support of the Funda\c{c}\~{a}o para a Ci\^{e}%
ncia e Tecnologia through the grants CERN/FP/123615/2011 and
CERN/FP/123618/2011.


\begin{references}

\bibitem{1} B. A. Bassett, S. Tsujikawa, and D. Wands, Rev. Mod. Phys. {\bf 78}, 537 (2006).

\bibitem{2} A. Maleknejada, M. M. Sheikh-Jabbaria, and J. Soda, Physics Reports {\bf  528},  161 (2013).

\bibitem{Li90} A. D. Linde, Particle physics and inflationary
cosmology, Harwood Academic Publishers, (1990).

\bibitem{Li98} A. R. Liddle, Phys. Rept. {\bf 307}, 53 (1998).

\bibitem{caldwell}
R. R. Caldwell, R. Dave, and P. J. Steinhardt,
Phys. Rev. Lett. {\bf 80}, 1582 (1998).

\bibitem{11} L. P. Chimento, A. S. Jakubi and D. Pavon, Phys. Rev. {\bf D  62}, 063508 (2000).


\bibitem{kessence}
T. Chiba, T. Okabe, and M. Yamaguchi,
Phys. Rev. {\bf D 62}, 023511 (2000);
%
C. Armendariz-Picon, V. F. Mukhanov, and P. J. Steinhardt,
Phys. Rev. Lett. {\bf 85}, 4438 (2000);
%
C. Armendariz-Picon, V. F. Mukhanov, and P. J. Steinhardt,
Phys. Rev. {\bf D 63}, 103510 (2001);
%
N. Arkani-Hamed, H. C. Cheng, M. A. Luty, and S. Mukohyama,
JHEP {\bf 0405}, 074 (2004);
%
F. Piazza and S. Tsujikawa,
JCAP {\bf 0407}, 004 (2004).

\bibitem{coupledDE}
C. Wetterich,
Astron. Astrophys. {\bf 301}, 321 (1995);
%
L.Amendola,
Phys. Rev. {\bf D 62}, 043511 (2000);
%
N. Dalal, K. Abazajian, E. E. Jenkins, and A. V. Manohar,
Phys. Rev. Lett. {\bf 87}, 141302 (2001);
%
W. Zimdahl, D. Pavon, and L. P. Chimento,
Phys. Lett. {\bf B 521}, 133 (2001);
%
S. del Campo, R. Herrera, G. Olivares, and D. Pavon,
Phys. Rev. {\bf D 74}, 023501 (2006);
%
H. Wei and S. N. Zhang,
Phys. Lett. {\bf B 644}, 7 (2007);
%
L. Amendola, G. C. Campos, and R. Rosenfeld,
Phys. Rev. {\bf D 75}, 083506 (2007);
%
Z. K. Guo, N. Ohta, and S. Tsujikawa,
Phys. Rev. {\bf D 76}, 023508 (2007);
%
G. Caldera-Cabral, R. Maartens, and L. A. Urena-Lopez,
Phys. Rev. {\bf D 79}, 063518 (2009);
%
B. Gumjudpai, T. Naskar, M. Sami, and S. Tsujikawa,
JCAP {\bf 0506}, 007 (2005);
%
L. Amendola and C. Quercellini,
Phys. Rev. {\bf D 68}, 023514 (2003).


\bibitem{DM_DE}
A. Y. Kamenshchik, U. Moschella, and V. Pasquier,
Phys. Lett. {\bf B 511}, 265 (2001);
%
M. C. Bento, O. Bertolami, and A. A. Sen,
Phys. Rev. {\bf D 66}, 043507 (2002);
%
R. J. Scherrer,
Phys. Rev. Lett. {\bf 93}, 011301 (2004).

\bibitem{Co94} E. J. Copeland, E. W. Kolb, A. R. Liddle and J. E.
Lidsey, Phys. Rev. {\bf D 49}, 1840 (1994).

\bibitem{Mi} F. E. Schunk and E. W. Mielke, Phys. Rev. {\bf D 50},
4794 (1994).

\bibitem{Caldwell:2005tm}
  R.~R.~Caldwell and E.~V.~Linder,
  Phys.\ Rev.\ Lett.\  {\bf 95}, 141301 (2005).

\bibitem{freezing}
B. Ratra and P. J. E. Peebles,
Phys. Rev. {\bf D 37} 3406 (1988);
%
I. Zlatev, L.M. Wang and P.J. Steinhardt,
Phys. Rev. Lett. {\bf 82}, 896 (1999);
%
P. Brax, J. Martin,
Phys. Lett. {\bf B 468}, 40 (1999);
%

  S.~Dutta and R.~J.~Scherrer,
  Phys.\ Lett. {\bf  B 704}, 265 (2011).


\bibitem{thawing}
R. Kallosh, J. Kratochvil, A. Linde, E.V. Linder, M. Shmakova,
JCAP {\bf 0310}, 015 (2003);
%
bibitem{Chiba:2012cb}
  T.~Chiba, A.~De Felice and S.~Tsujikawa,
  Phys.\ Rev.\ {\bf D  87}, 083505 (2013);
T.~G. Clemson and A.~R. Liddle,
  Mon.\ Not.\ Roy.\ Astron.\ Soc.\  {\bf 395}, 1585 (2009);
S.~del Campo, V.~H.~Cardenas and R.~Herrera,
  Phys.\ Lett.\ {\bf B  694}, 279 (2011);
N.~C.~Devi and A.~A. Sen,
  Mon.\ Not.\ Roy.\ Astron.\ Soc.\  {\bf 413}, 2371 (2011).

\bibitem{Barreiro:1999zs}
  T.~Barreiro, E.~J.~Copeland and N.~J.~Nunes,
  Phys.\ Rev.\ {\bf D  61}, 127301 (2000).

\bibitem{general_pot}
V. Sahniand, L.M. Wang,
Phys. Rev. {\bf D 62}, 103517 (2000).

\bibitem{P1} P. A. R. Ade et al., Planck 2013 results. I, arXiv: 1303.5062
[astro-ph) (2013).

\bibitem{P2} P. A. R. Ade et al., Planck 2013 results. XXVI, arXiv:1303.5086 (astro-ph) (2013).

\bibitem{3} J. Elliston, D. J. Mulryne, and R. Tavakol, Phys. Rev. {\bf D 88}, 063533 (2013).

\bibitem{4} K. Nakayama, F. Takahashi, and T. T. Yanagida, Phys. Lett. {\bf B 725}, 111 (2013).

\bibitem{acc} D. H. Weinberg, M. J.  Mortonson, D. J. Eisenstein, C. Hirata, A. G.  Riess, and E. Rozo, Physics Reports {\bf 530}, 87 (2013).

\bibitem{PeRa03} P. J. E. Peebles and B. Ratra, Rev. Mod. Phys.
{\bf 75}, 559 (2003).

\bibitem{Pa03} T. Padmanabhan, Phys. Repts. {\bf 380}, 235 (2003).

\bibitem{8} R. Caldwell, R. Dave and P. J. Steinhardt, Phys. Rev. Lett. {\bf 80}, 1582 (1998).

\bibitem{Tsu}  S. Tsujikawa,  Class. Quant. Grav. {\bf 30},  214003 (2013).

\bibitem{Harko:2012za}
  T.~Harko and F.~S.~N.~Lobo,
  Phys.\ Rev.\ {\bf D  87},  044018 (2013).

\bibitem{Harko:2013wsa}
  T.~Harko and F.~S.~N.~Lobo,
  JCAP {\bf 1307}, 036 (2013).

  \bibitem{gal1} A. Nicolis, R. Rattazzi, and E. Trincherini, Phys. Rev. {\bf D 79}, 064036 (2009).


  \bibitem{gal2}  C. Defayet, G. Esposito-Farese and A. Vikman, Phys. Rev. {\bf D 79}, 084003 (2009);
 C. de Rham and A. J. Tolley, JCAP {\bf 1005}, 015 (2010);
G. Goon, K. Hinterbichler and M. Trodden, Phys. Rev. Lett. {\bf 106}, 231102 (2011);
 G. Goon, K. Hinterbichler and M. Trodden, JCAP {\bf 1107}, 017 (2011).

 \bibitem{gal3}
K. Kamada, T. Kobayashi, M. Yamaguchi and J. ’i. Yokoyama, Phys. Rev. {\bf D 83}, 083515
(2011).

\bibitem{gal4} T. Kobayashi, M. Yamaguchi and J. ’i. Yokoyama, Prog. Theor. Phys. {\bf 126},  511 (2011).

\bibitem{Rub} V. A. Rubakov, arXiv:1401.4024 (2014).

\bibitem{phan1}  R. R. Caldwell, Phys. Lett. {\bf B. 545}, 23 (2002).

\bibitem{phan2} S. M. Carroll, M. Hoffman, and M. Trodden,
Phys. Rev. {\bf D 68}, 023509 (2003);
%
P. Singh, M. Sami, and N. Dadhich,
Phys. Rev. {\bf D 68}, 023522 (2003);
%
M. Sami and A. Toporensky,
Mod. Phys. Lett. {\bf A 19}, 1509 (2004);
%
J. M. Cline, S. Jeon, and G. D. Moore,
Phys. Rev. {\bf D 70}, 043543 (2004); E.~Elizalde, S.~Nojiri and S.~D.~Odintsov, Phys. Rev. {\bf D 70}, 043539 (2004); E.~Elizalde, S.~Nojiri, S.~D.~Odintsov, D.~Saez-Gomez and V.~Faraoni, Phys.\ Rev. {\bf D 77},  106005 (2008).

\bibitem{phan3} A. Yu. Kamenshchik, Class. Quantum Grav. {\bf 30}, 173001 (2013).

\bibitem{phan4} U. Alam, V. Sahni, T. D. Saini, and A. A. Starobinsky, Mon. Not. Roy. Astron. Soc. {\bf 354}, 275 (2004).

\bibitem{Mus} A. G. Muslimov, Class. Quantum Grav. {\bf 7}, 231 (1990).

  \bibitem{Sal} D. S. Salopek and J. R. Bond, Phys. Rev. {\bf D 42}, 3936 (1990).

  \bibitem{Giambi} R. Giambo, Class. Quantum Grav. {\bf 22},  2295 (2005); R. Giambo, F. Giannoni, and G. Magli, J. Math. Phys. {\bf 49},  042504 (2008); R. Giambo, F. Giannoni, and G. Magli, Gen. Rel. Grav {\bf 41},  2130 (2009); R. Giambo and A. Stimilli, J. Geom. Phys. {\bf  59},   400 (2009).

      \bibitem{Kis} V. V. Kiselev, JCAP {\bf 0801},  019 (2008).

\bibitem{Lid0} J. E Lidsey, A. R Liddle, E. W Kolb, E. J. Copeland, T. Barreiro, and M. Abney, Rev. Mod. Phys. {\bf 69}, 373 (1997).

\bibitem{Kam1}  A. Y. Kamenshchik, A. Tronconi, and G. Venturi, Phys. Lett. {\bf B 702}, 191 (2011).
\bibitem{Kam2} A. Yu. Kamenshchik, A. Tronconi, G. Venturi, S. Yu. Vernov, Phys.
Rev. {\bf D 87}, 063503 (2013).

\bibitem{Kam3} A. Yu. Kamenshchik, E. O. Pozdeeva, A. Tronconi, G. Venturi, and S. Yu. Vernov, 	arXiv:1312.3540 (2013).

\bibitem{Nunes:2000yc}
  A.~Nunes and J.~P.~Mimoso,
  Phys.\ Lett. {\bf B 488}, 423 (2000).

\bibitem{Mimoso:2005bv}
  J.~P.~Mimoso, A.~Nunes and D.~Pavon,
  Phys.\ Rev. {\bf D 73}, 023502 (2006).

  \bibitem{Charters:2009ku}
  T.~Charters and J.~P.~Mimoso,
  JCAP {\bf 1008}, 022 (2010).

\bibitem{Lid1} J. E. Lidsey, Phys. Rev. {\bf D 86}, 123523 (2012).

\bibitem{Lid2} J. E. Lidsey, 	arXiv:1309.7181 (2013).

\bibitem{Ber} A. E. Bernardini and O. Bertolami,  Annals of Physics {\bf 338}, 1 (2013).

\bibitem{Fre1} P. Fre, A. Sagnotti, and A. S. Sorin, to appear in Nucl. Phys. B, 	arXiv:1307.1910 (2013).

\bibitem{Fre2} P. Fre, A.S. Sorin, and M. Trigiante, 	arXiv:1310.5340 (2013).

\bibitem{exp1} J.D. Barrow, Phys. Lett. {\bf B 187}, 12 (1987); A. B. Burd and J. D. Barrow, Nucl. Phys. {\bf B 308}, 929 (1988); L. P. Chimento,
 Class. Quant. Grav. {\bf 15}, 965 (1998); J. G. Russo, Phys. Lett. {\bf B 600}, 185 (2004); C. Rubano, P. Scudellaro, E. Piedipalumbo, S. Capozziello, and M. Capone, Phys. Rev. {\bf D 69}, 103510 (2004).

\bibitem{K1}A. Andrianov, F. Cannata, and A. Yu. Kamenshchik, JCAP  {\bf 10},  004 (2011).

 \bibitem{K2} A. A. Andrianov, F. Cannata, and A. Y. Kamenshchik, Phys. Rev. {\bf D 86},  107303 (2012). 	
\bibitem{Cui} W.-P. Cui, Y. Zhang, and Z.-W. Fu, Research in Astronomy and Astrophysics {\bf 13},  629 (2013).

\bibitem{CaMaPeFr85} C.G. Callan, E.J. Martinec, M.J. Perry and D.Friedan, Nucl. Phys. {\bf B 262}, 593 (1985).

\bibitem{CaCaMu93} B. de Carlos, J. A. Casas and C. Munoz, Nucl. Phys. {\bf B 399}, 623 (1993).

\bibitem{Sch} H.-J. Schmidt, Astron. Nachr. {\bf 311}, 165 (1990).

\bibitem{Sch1} H.-J. Schmidt and A. A. Starobinsky, Class. Quantum Grav. {\bf 7}, 1163 (1990).

\bibitem{Sch2} C.-M. Chen, T. Harko, and M. K. Mak, Phys. Rev. {\bf D 62},  124016 (2000).

\bibitem{pow1} F. Lucchin and S. Matarrese S, Phys. Rev. {\bf D 32} 1316 (1985); J. J. Halliwell, Phys. Lett. {\bf B 185}, 341 (1987); A. B. Burd and J. D. Barrow, Nucl. Phys. {\bf B 308}, 929 (1988); G. F. R. Ellis and M. S. Madsen, Class. Quant. Grav.
{\bf 8}, 667 (1991); A. R. Liddle and R. J. Scherrer,  Phys. Rev. {\bf D 59}, 023509 (1999); V. Gorini, A. Y. Kamenshchik, U. Moschella and V. Pasquier,  Phys. Rev. {\bf D 69}, 123512 (2004).

\bibitem{pow} M. S. Turner, Phys. Rev. {\bf D 28}, 1243 (1983); S. Tsujikawa, Phys. Rev. {\bf D 62},  043512 (2000);   A. de la Macorra and G. German, Phys. Lett. {\bf B 549}, 1 (2002); J. Martin and C. Ringeval, Phys. Rev. {\bf D 82}, 023511 (2010); M.-L. Tong, Class. Quantum Grav. {\bf 30}, 055013 (2013).

\bibitem{PoZa} A. D. Polyanin and V. F. Zaitsev, Handbook of exact solutions for ordinary differential equations, CRC Press, Boca Raton, New York, London, Tokyo (1995).

\end{references}
\end{document}